\documentclass[aps,pra,superscriptaddress,noshowpacs]{revtex4}

\usepackage[utf8]{inputenc}
\usepackage[sort&compress]{natbib}
\usepackage{amsfonts,amsmath,amssymb}
\usepackage{graphicx}
\usepackage[english]{babel}
\usepackage{newlfont}
\usepackage[T1]{fontenc}
\usepackage{wrapfig}
\usepackage{float}

\DeclareMathOperator{\DBCS}{\Delta_{BCS}}

\newcommand{\bra}[1]{\left\langle #1 \right|}
\newcommand{\ket}[1]{\left| #1 \right\rangle}

\newcommand{\vev}[1]{\left\langle #1 \right\rangle}

\begin{document}

\title{Relative Phase and Josephson Dynamics 
between Weakly Coupled Richardson Models}

\author{Francesco Buccheri}
\affiliation{SISSA and INFN, Sezione di Trieste, Via Bonomea 265, 
I-34136 Trieste, Italy}
\affiliation{Wigner Research Centre for Physics, Konkoly Thege Miklós út 29-33, 1121 Budapest, Hungary}
\affiliation{MTA-BME "Momentum" Statistical Field Theory Research Group,
Budafoki ut 8, H-1111 Budapest, Hungary}

\author{Andrea Trombettoni}
\affiliation{CNR-IOM DEMOCRITOS Simulation Center, Via Bonomea 265, 
I-34136 Trieste, Italy}
\affiliation{SISSA and INFN, Sezione di Trieste, Via Bonomea 265, 
I-34136 Trieste, Italy}


\begin{abstract}

We consider two weakly coupled Richardson models to study 
the formation of a relative phase and the Josephson dynamics between 
two mesoscopic attractively interacting fermionic systems: our results 
apply to superconducting properties of coupled ultrasmall metallic grains 
and to Cooper-pairing superfluidity in neutral systems with a finite number 
of fermions. We discuss how a definite relative phase 
between the two systems emerges and how it can be 
conveniently extracted 
from the many-body wavefunction: we find that a definite 
relative phase difference emerges even for very small numbers of pairs 
($\sim 10$). 
The Josephson dynamics and the current-phase characteristics are then 
investigated, showing that the critical current has a maximum at the BCS-BEC 
crossover. For the considered initial conditions 
a two-state model gives a good description of the dynamics 
and of the current-phase characteristics.
\end{abstract}

\maketitle

\section{Introduction}\label{sec:introduction}

A major issue in mesoscopic physics is the study of the 
sample sizes for which macroscopic properties emerge in finite systems 
\cite{book}. 
A typical context for such a study 
is provided by systems exhibiting quantum coherence, 
e.g. superconductivity or superfluidity, in 
samples with restricted size or number of particles: the general 
question is then to determine when and how quantum 
coherence takes place. The subsequent study 
is relevant in a number of situations, ranging from the investigation 
of superfluidity of $He$ droplets \cite{droplets} to Bose-Einstein condensation 
in atomic gases with small number of particles \cite{L06}. 

A prototypical example of these studies of quantum coherence in mesoscopic 
interacting systems is given by the investigation of the limit size 
of a metallic grain needed for 
the occurrence of superconductivity \cite{A59}: 
this and related questions are conveniently studied by using 
the Richardson model (RM) 
\cite{DR01,DPS04}. The RM describes a system of attractively 
interacting fermions
and is paradigmatic in characterizing pairing in systems with a 
finite number of fermions \cite{DPS04}: its relevance is 
also due to the remarkable feature of being exactly solvable \cite{R63} and 
to the fact that it is possible 
to derive the thermodynamic limit of its exact solution and show that 
it precisely coincides with the BCS solution \cite{R77,G95}. 

The RM has been first studied in 
the context of nuclear physics \cite{R63,RS64}, where the attraction 
leading to the pairing is due to the short-range nature 
of the effective nucleon-nucleon interaction \cite{DHJ03}. 
It was subsequently shown to be 
deeply connected with the exactly solvable Gaudin magnets \cite{G75},
through the relation between the respective integrals of motion 
\cite{CRS97}. Using such relation, the Richardson model was then extended 
to more general classes of exactly solvable pairing-like models 
\cite{ADLO01,DES01,AFF01,DP02,ZLMG02}.

The RM is particularly relevant 
for the study of finite-size scaling effects in the BCS theory of 
superconductivity 
\cite{BD98,DS99,DS00,RSD02,YBA03,YBA05,FCC08,FCC10,AO12}. The 
reason is that the classic BCS approach to superconductivity 
\cite{BCS57} in the presence of a pairing interaction violates 
particle number conservation \cite{L06}: 
number fluctuations are negligible in the thermodynamic limit, 
but important for small number of particles \cite{DPS04}. 
For this reason, the RM is used in the analysis of 
ultrasmall metallic and superconducting nanograins 
\cite{DR01}: experiments on such systems
are actually performed at a fixed number of electrons \cite{RBT95} 
due to their large charging energy. The Richardson model 
was successfully used in clarifying many features of 
the tunneling spectra of $Al$ nanograins 
\cite{RBT95,BRT96,BD99}, where, for instance, the spectroscopic gap between 
grains with an odd or even number of electrons was explained with 
the existence of pairing correlations among these \cite{MFF98}. 

It is a known general fact that 
when two superconducting or superfluid systems are weakly 
coupled a supercurrent flows between the two systems, with the current 
depending on the relative phase between the two superconducting or superfluid 
systems \cite{J62,BP82}. The importance of this Josephson effect stems from the 
fact that it describes coherent tunneling between 
superfluid/superconducting systems, and this description is in most cases 
independent 
on the details of the microscopic description of the uncoupled systems and 
of the concrete physical realization of the weak link between them. 
In this paper we intend to investigate how a definite 
relative phase emerges between two mesoscopic 
finite-size attractively interacting systems modelled by RMs and 
how it is possible to extract it from the time-dependent 
many-body wavefunction: we find that this happens
even for very small total number of pairs ($\sim 8-10$). 
This occurs when the ``bulk'' interaction (i.e., the paring interaction
of the uncoupled systems) is such that the equilibrium properties of 
the uncoupled models are rather well approximated by the large-$N$ BCS theory. 
Once the phase is formed and extracted from the many-body wavefunction, 
it is then possible to determine the current-phase portrait and study 
the Josephson effect in such mesoscopic weakly coupled fermionic systems. 

Our results can be primarily applied to weakly 
coupled ultrasmall metallic grains \cite{DR01}, 
but they could be also useful in connection with 
cold atom experimental setups in which the trapping potential 
contains a small number 
of fermions (like \cite{SZLOWJ11}) and such traps are 
set at a distance that allows tunneling: this would be the atomic 
counterpart of superconducting ultrasmall grains coupled by tunneling terms. The fate of the Josephson effect between small 
superconducting grains was investigated to some extent in \cite{GSD04}, studying 
the dependence of the Josephson energy as function of the level spacing and 
focusing on a parameter regime where
 the notion of a superconducting phase variable is not valid. 

Another application of the Richardson model 
is to the study of the BCS-BEC crossover
in finite size fermionic systems \cite{OD05}. The BCS-BEC crossover 
is a subject which  
has been thoroughly investigated, also in connection to experimental 
realizations with ultracold fermions \cite{CSTL05,GPS08,Z12}. 
Increasing the (bare) attractive interaction among fermions, the chemical 
potential decreases with respect to the non-interacting Fermi energy value, 
so that a crossover between a BCS state, 
characterized by loosely correlated, widely overlapping Cooper pairs, 
to a Bose-Einstein condensate (BEC), in which pairs are tightly bound and 
minimally overlapping, can be identified \cite{CSTL05,GPS08,Z12}. 
Within the formalism of the Richardson model, 
the corresponding finite-$N$ version of the BCS-BEC crossover \cite{OD05}, 
as well as the Josephson effect, can be studied.

In this paper we numerically investigate the 
Josephson dynamics of two weakly coupled 
Richardson Hamiltonians. Our motivation for such an investigation 
is three-fold. 
From one side we are interested in characterizing 
the superfluid behavior of the system at finite number of particles, 
with regard to its phase coherence, and in investigating 
for which values of the number of particles 
a definite relative phase between the two systems is formed:  
we find that the system behaves coherently 
even for a rather small total number of pairs (as low as $\sim 8 - 10$)
We introduce and discuss a way to extract from the many-body wavefunction 
the relative phase and its variance, so to quantify in a precise manner 
whether a well definite relative phase emerge.

When the relative phase is well defined, we are then 
interested in understanding and characterizing  
the effects of the pairing interaction coefficient $g$ (giving rise 
in the uncoupled models to the BCS-BEC crossover) on the coupled dynamics 
while varying the pairing interaction coefficient. We will be mostly 
interested to values of coupling $g$ such that the uncoupled models 
have level occupation amplitudes close to the large-$N$ results. 

Finally, our work aims at providing the exact Josephson 
dynamics between two weakly coupled Fermi systems with small number of 
fermions across the BCS-BEC crossover. Theoretical studies of 
tunneling of ultracold fermions across 
the BCS-BEC crossover recently appeared 
\cite{TD05,SPS07,SMT08,WODPS08,AST09,WDPPS09,SPS10,WDPS11,IFT12}. 
In \cite{SPS07} the tunneling across a barrier potential was studied 
by solving numerically the Bogoliubov-de Gennes equations at zero temperature: 
the Josephson current was found to be enhanced around the unitary limit. 
For vanishing barriers (i.e. large coupling between the two Fermi systems), 
the critical current approaches the Landau limiting value \cite{SPS07}. 
Results obtained from the numerical solution of the Bogoliubov-de Gennes 
equations were compared with the analytical predictions derived 
from a hydrodynamic scheme, in the local density approximation 
\cite{WDPPS09}: whenever such approximation is valid 
(small and intermediate barriers), good agreement was found. 
In general, it is instead more difficult to obtain solutions 
of the Bogoliubov-de Gennes equations for very large barriers \cite{SPS10} 
i.e., when the coupling between the two Fermi systems is weak. 
Furthermore, one would also like to explore the exact tunneling dynamics 
and eventually compare it with the time-dependent solution of 
the Bogoliubov-de Gennes equations, 
which has been successfully 
used to study the dynamics of soliton solutions 
in trapped superfluid Fermi gases \cite{SDPS11}.
 
The model which is studied in the present paper, 
although necessarily restricted to small number of particles, exploits 
the integrability of the two uncoupled Richardson systems and 
allows to compute the exact dynamics when a state with non-vanishing 
initial number imbalance and/or relative phase is prepared, offering the 
opportunity to extract the dynamical phase portrait. Another advantage 
is that it is possible to investigate, in a simplified setting, 
how the Josephson energy depends on the interaction and the tunneling strength: 
we find that the Josephson energy 
has a maximum around the unitary limit, 
in agreement with results in literature obtained at $T=0$ 
in the large-$N$ limit for small and intermediate barriers \cite{SPS07,WODPS08}.

The plan of the paper is the following: in Section \ref{sec:Richardson} 
we review the main properties of a single (i.e., uncoupled) 
Richardson model. The model with two coupled Richardson Hamiltonians is 
introduced in Section \ref{sec:coupled}, where we also discuss 
the main properties of the spectrum. The Josephson 
dynamics is studied in Sections \ref{sec:phase}-\ref{sec:dyn}: in Section 
\ref{sec:phase} we introduce the considered 
initial states for the dynamics and 
we discuss the emergence of a definite relative phase among 
the two Richardson systems. In Section \ref{sec:dyn} we discuss the 
dynamical phase portrait, plotting the trajectories in the space of the 
relative phase and the population imbalance, and we present our results 
for the critical current as a function of the coupling. 
We draw our conclusions in Section \ref{sec:concl}.

\section{The Richardson model}\label{sec:Richardson}

The Richardson Hamiltonian is written in terms of the operators 
$c_{\alpha\sigma}$ destroying fermionic particles in 
the energy levels $\alpha=1,\ldots,N$ with spin $\sigma=\uparrow,\downarrow$:
\begin{equation}\label{RichardsonHamiltonian}
H = \sum_{\alpha=1}^{N} \varepsilon_\alpha 
\left(c^{\dagger}_{\alpha\uparrow}c_{\alpha\uparrow}+c^{\dagger}_{\alpha\downarrow}c_{\alpha\downarrow}\right)
- 2 g \sum_{\alpha,\beta=1}^{N} 
c^{\dagger}_{\alpha\uparrow}c^{\dagger}_{\alpha\downarrow} 
c_{\beta\downarrow}c_{\beta\uparrow}.
\end{equation}
In Eq. (\ref{RichardsonHamiltonian}) the $\varepsilon_\alpha$ are 
the single-particles energies of the $N$ levels, 
$g$ is an interaction coefficient with the dimensions of an energy: 
in the following $g$ is assumed to be positive 
(corresponding to attraction among 
fermions) and it models the matrix element of the scattering among Cooper 
pairs of spin-reversed fermions. The model is integrable for any choice
 of the set of energies $\varepsilon_\alpha$ - in the following 
we will consider them to equally spaced, according 
\begin{equation}\label{eq-spac}
\varepsilon_\alpha \equiv  \alpha d,
\end{equation}
where $\alpha=1,\ldots,N$ and $d$ is the level spacing: 
this is indeed the choice usually done 
in order to recover the BCS physics in the thermodynamic limit 
(see more details below) \cite{R77,DPS04}. 

The Richardson Hamiltonian (\ref{RichardsonHamiltonian}) conserves 
the number of fermions and, separately, 
of fermion pairs (doubly occupied levels). 
An essential feature of the spectrum is the so-called 
``blocking'' effect \cite{DPS04}: 
the states which are singly occupied, i.e. 
those in which there is only one electron with either $\uparrow$ 
or $\downarrow$ spin, 
are unaffected by the interaction and the net effect 
arising from their presence 
is that of "blocking" the level by preventing the 
scattering of the other pairs on it. 
The full Hilbert space is then divided into sectors with a given number of 
unpaired fermions and in each of these subspaces 
the Hamiltonian (\ref{RichardsonHamiltonian}) only couples the doubly occupied 
("unblocked") levels among them, 
while leaving singly-occupied levels effectively decoupled from the dynamics. 
Denoting the number of pairs by $M$, 
it is customary to write a reduced Hamiltonian for the
 $N_f=2M$ paired fermions in the $N$ unblocked levels as: 
\begin{equation}\label{RichardsonHamiltonian_bosons}
H = 2 \sum_{\alpha=1}^{N} \varepsilon_\alpha b^\dagger_\alpha b_\alpha
- 2g \sum_{\alpha,\beta=1}^{N} b^\dagger_{\alpha}b_{\beta}, 
\end{equation}
where we introduced the (hardcore) pair creation and annihilation operators 
\begin{equation}\label{fermionsToSpins}
b^\dagger_\alpha = c^{\dagger}_{\alpha\uparrow}c^{\dagger}_{\alpha\downarrow}  \;, \quad 
b_\alpha  = c_{\alpha\downarrow} c_{\alpha\uparrow}.
\end{equation}
Notice that the Hilbert spaces on which the Hamiltonian 
(\ref{RichardsonHamiltonian_bosons}) 
acts are subspaces of the full space of 
(\ref{RichardsonHamiltonian}), characterized by given
 configurations of blocked levels.

An Hamiltonian equivalent to (\ref{RichardsonHamiltonian_bosons}) can be
written by introducing the Anderson 
pseudospin-$1/2$ operators \cite{A58}: 
{$S_\alpha^-=b_\alpha$, $S_\alpha^+=b_\alpha^{\dagger}$, 
$2S_\alpha^z = c^{\dagger}_{\alpha\uparrow} c_{\alpha\uparrow} + c^{\dagger}_{\alpha\downarrow}c_{\alpha\downarrow}-1$}. In terms 
of the $su(2)$ algebra generators, up to a constant, one has
\begin{equation}\label{RichardsonHamiltonian_spins}
H = 2 \sum_{\alpha=1}^{N} \varepsilon_\alpha S^z_\alpha
- 2g \sum_{\alpha,\beta=1}^{N} S^+_{\alpha}S^-_{\beta}.
\end{equation}
Explicit solutions of the dynamics generated by the Hamiltonians 
(\ref{RichardsonHamiltonian_bosons}) and (\ref{RichardsonHamiltonian_spins}) has been presented and discussed 
in \cite{YAKE05}.

The exact (not normalized) eigenstates of (\ref{RichardsonHamiltonian_bosons}) 
are constructed by applying a set of generalized creation operators $\tilde B$ 
on the reference state $\ket{0}$ as follows:
\begin{equation}\label{BetheState}
\ket{\{w\}}=\prod_{j=1}^M \tilde B(w_j) \ket{0},
\end{equation}
where the reference state is the one in which no hardcore bosons are present:
\begin{equation}\label{0}
b_\alpha \ket{0} = 0 \;\qquad (\alpha=1,2,\ldots, N).
\end{equation}
The explicit form of the creation operators is
\begin{equation} \label{RichardsonB}
\tilde B(w)=\sum_{\alpha=1}^{N}\frac{b^\dagger_\alpha}{w-\varepsilon_\alpha}
\end{equation}
and the set of complex number $\{w_j\}$ (referred to as rapidities) 
satisfies the set of algebraic equations
\begin{equation}\label{RichardsonEquations}
\frac{1}{g}+\sum_{\alpha=1}^{N}\frac{1}{w_j-\varepsilon_\alpha} - \sum_{k \ne j}^{N} \frac{2}{w_j-w_k} = 0 \qquad (j=1,\ldots,M).
\end{equation}

The number of rapidities corresponds to the number of Cooper pairs 
in the state and the action of the operator (\ref{RichardsonB})
 is that of creating a boson, with a given amplitude on each level $\alpha$,
 as results from the interaction with all the other Cooper pairs, which in 
 turn is encoded in the system (\ref{RichardsonEquations}).
 In the limit $g \to 0$ all the roots of the Richardson equations (\ref{RichardsonEquations}) 
are real and coincide with a given subset of fields, 
so that each boson is localized on a definite energy level.
On the other hand, when $g$ is moved to nonzero values, roots
can be present in complex conjugated pairs. 
In particular, for $g \to 0$, the ground state for a given number $M$ of pairs 
is the one in which the lowest $M$ levels are filled;
 in the strong coupling limit $g\to\infty$ all the roots of this state
 come in complex pairs (except for the most negative one,
 when $M$ is odd)  and their absolute value diverges. 
The BCS equations can be obtained from this solution in 
the limit $N\to\infty$ while keeping constant filling $M/N$, 
energy range $Nd$ and effective coupling strength $g N$. 
In this limit, the root configuration associated 
to the ground state assumes the shape of an arch in the complex 
plane, whose extrema are at 
\begin{equation}
\mu\pm i \DBCS.
\label{arch}
\end{equation}
As shown in \cite{R77,G75,RSD02,DPS04}, 
the link of the finite-$N$ results with the large-$N$ BCS theory 
is provided by the fact that $\DBCS$ and 
$\mu$ satisfy in the scaling limit previously defined the BCS equations
\begin{equation}\label{BCS-number-equation}
2M=\sum_{\alpha} \left(1-\frac{\varepsilon_\alpha-\mu}{\sqrt{(\varepsilon_\alpha-\mu)^2+\Delta_{BCS}^2}}\right)
\end{equation}
and
\begin{equation}\label{BCS-gap-equation}
\frac{1}{g}=\sum_{\alpha} \frac{1}{\sqrt{(\varepsilon_\alpha-\mu)^2+\Delta_{BCS}^2}}.
\end{equation}

The Richardson mode is integrable by means of algebraic Bethe ansatz \cite{CRS97,DP02}:
 not only the spectrum and the eigenstates, 
but also matrix elements \cite{LZMG03} and 
correlation functions \cite{S99,AO02,ZLMG02,FCC08,FCC10,AO12} 
are exactly computable. In particular, 
given two states $\ket{ \{v\}}$ and $\ket{ \{w\}}$ 
defined as in (\ref{BetheState}) with $M$ rapidities, one can 
make use of
\begin{eqnarray}\label{Sz-Rich}
\left\langle \{w\} \right| b_\alpha^\dagger b_\alpha-\frac{1}{2} \left| \{v\} \right\rangle &=&
-\prod_{l=1}^M\frac{w_l-\varepsilon_\alpha}{v_l-\varepsilon_\alpha}\frac{1}{\prod_{k>j}^M(v_k-v_j)(w_j-w_k)}\det\left[ \tilde{H} - 2 \tilde{P}_\alpha \right],
\end{eqnarray}
where $\tilde{H}$ is a $M\times M$ matrix defined as
\begin{equation}\label{tildeH}
  \tilde H_{j,k} =   \frac{\prod_{l=1}^M (w_l-v_k)}{(w_j-v_k)^2}
\left( \frac{1}{g}-\sum_{\alpha=1}^{N}\frac{1}{v_k-\varepsilon_\alpha}+\sum_{l\ne j}\frac{2}{v_k-w_l}\right).
\end{equation}
$\tilde P$ is given by
\begin{equation}\label{tildePmat}
\left[\tilde P_\alpha\right]_{j,k}
 = \frac{\prod_{l\ne k}(v_l-v_k)}{(w_j-h_\alpha)};
\end{equation}
Moreover, the following relation will be also used: 
\begin{eqnarray}\label{SpSmFF-Rich}
\left\langle \{v\} \right| b_\alpha \left| \{w\} \right\rangle = \left\langle \{w\} \right| b_\alpha^\dagger \left| \{v\} \right\rangle 
&=& \frac{\prod_{l=1}^M\left(w_l-h_\alpha\right)}{\prod_{l=1}^{M-1}\left(v_l-h_\alpha\right)}
\frac{\det \tilde H^-}{\prod_{j<k}\left(v_k-v_j\right)\left(w_j-w_k\right)}, 
\end{eqnarray}
in which the state $\ket{ \{v\}}$ has now $M-1$ rapidities and 
the $M \times M$ matrix 
$\tilde H^-$ is defined as:
\begin{equation}\label{tildeH-}
\tilde H^-_{j,k} =\left\{\begin{array}{lc}
                    \frac{\prod_{l=1}^M (w_l-v_k)}{(w_j-v_k)^2}
\left( \frac{1}{g}-\sum_{\alpha=1}^{N}\frac{1}{v_k-\varepsilon_\alpha}+\sum_{l\ne j}\frac{2}{v_k-w_l}\right)
		    & k<M \\
		     \frac{1}{(w_j-h_\alpha)^2} & k=M.
                   \end{array}
\right.
\end{equation}

\subsection{BCS-BEC crossover in the Richardson model}\label{sec:BEC-BCS}

The Richardson model exhibits two types of crossover behavior: 
first, the crossover from bulk to few fermions, i.e. from large to small 
$N$ \cite{R77}. In this case 
the Richardson model is used to study the corrections to the large-$N$ BCS 
theory \cite{G95} and in general how the physical quantities are modified when 
the number $N$ is not large and the energy scale $d$ explicitly plays  
a role. Since we will numerically study the spectrum 
and the dynamics of coupled Richardson models, 
the size of the considered 
systems will be necessarily finite. Moreover, we will need
to solve the equations (\ref{RichardsonEquations}) to determine the 
eigenstates, which is best done when the spacing $d$ of the levels is kept 
finite while increasing the number of levels. It is then 
convenient to define an 
intensive Richardson gap \cite{FCC08}, which is related to the 
BCS gap $\DBCS$ by
\begin{equation}\label{Gap-RvsBCS}
\DBCS = N \Delta
\end{equation}
in which the quantity $\DBCS$ can be extracted 
from the ground state solution of the Richardson 
equations (\ref{RichardsonEquations}): it is found that 
$\DBCS \approx Ng$ \cite{DPS04}. 
In the large-$N$ limit, the correlation functions are given by 
\begin{equation}\label{uandvfrombcs}
\vev{b_\alpha^\dagger b_\alpha} = v^2_\alpha \;,\quad 
\vev{b_\alpha b_\alpha^\dagger} = u^2_\alpha \;,\quad 
\vev{b_\alpha^\dagger b_\beta} = u_\alpha v_\alpha u_\beta v_\beta \;\;\quad 
(\alpha\ne\beta)
\end{equation}
where the $u,v$'s enter the BCS variational ansatz for the ground-state 
$\left|GS\right\rangle = \prod_\alpha (u_\alpha+v_\alpha b^\dagger_\alpha) 
\left|0\right\rangle$ \cite{L06}. The study of the comparison between the 
correlation functions given by (\ref{uandvfrombcs}) with those directly 
from the Richardson model shows that increasing $g$ the agreement becomes better 
and better: e.g., as one can sees from Fig. 5 of \cite{FCC08} one has a 
rather good agreement already for $N \sim 10$ for $g \gtrsim 0.3$. 
We can then conclude that for values of $N$ considered 
in the rest of the paper 
one has for uncoupled systems a rather good agreement with large-$N$ results. 

The behavior of $\mu$, the real value of the extremes (\ref{arch}) 
of the arch formed by the 
Bethe roots in the complex plane 
for large values of $N$ and $g$, depends in general on the filling, 
and it is $\mu \propto -g$ for fixed values of the initial population
 imbalance, below half filling. 
An important point to be stressed is that in the thermodynamic 
limit the quantity $\mu$, as defined from the root configuration,
tends to the chemical potential obtained for attractively 
interacting fermions in the BCS-BEC crossover \cite{OD05}.

\begin{figure}
\begin{center}
\includegraphics[width=0.49\textwidth]{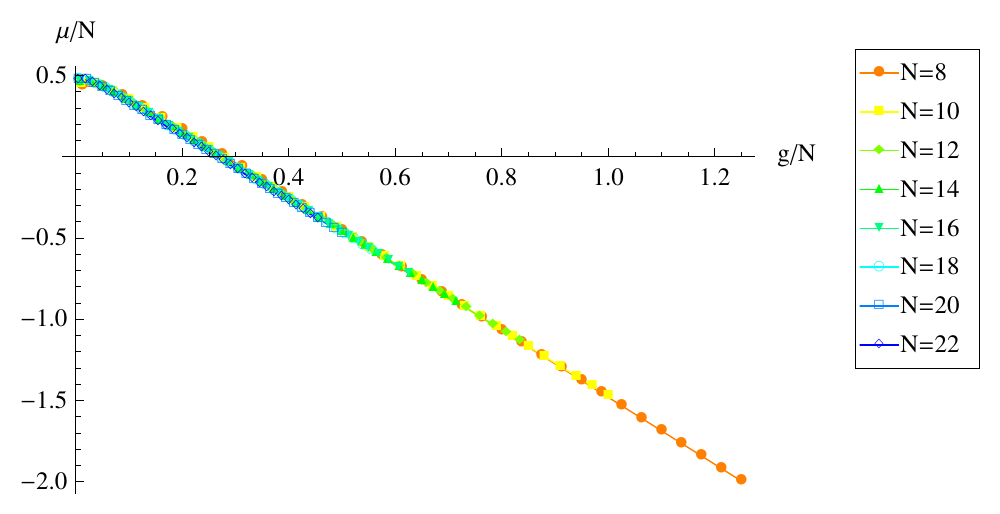}
\caption{Chemical potential $\mu$ per level versus $g/N$, as computed from 
Eqs. (\ref{BCS-number-equation})-(\ref{BCS-gap-equation}) for 
$M=N/2-1$. Here and in the captions of the following figures 
the pairing coefficient $g$ and the energies are expressed 
in units of $d$.}\label{fig:scaling}
\end{center}
\end{figure}

We then can argue that the other crossover taking place in the Richardson model is the BCS-BEC one: 
for large $N$ the parameters $\DBCS$ and $\mu$ satisfy 
Eqs. (\ref{BCS-number-equation})-(\ref{BCS-gap-equation}). Since the chemical 
potential changes sign for $g$ larger than a critical value, therefore 
a BCS-BEC crossover takes place \cite{Z12}. 
A description of the BCS-BEC crossover in the framework of the 
integrable Richardson model was given in \cite{OD05}, where the model 
(\ref{RichardsonHamiltonian_bosons}) was considered in the thermodynamic 
limit and it was argued there that root configurations at 
strong enough coupling can be used to identify the boundaries of the crossover. 
In Figure \ref{fig:scaling} we plot $\mu$ as a function 
of $g$ for different values of $N$ as determined from 
Eq. (\ref{BCS-number-equation}): 
for the considered values of $N$ one sees that 
$\mu$ changes sign for $g/N \sim 0.25d$ for $M$ close to $N/2$ 
(notice that exactly at half-filling 
$\mu$ does not change sign). Note that, whenever $M < N/2$, 
the chemical potential becomes more and more negative 
while increasing $g$: at some point, it crosses the real axis 
to negative values, signaling the crossover. Notice that 
in the BCS scaling 
\cite{DPS04}, in which the level spacing goes to zero as the inverse 
of the size, the crossing point tends to a finite value of $g$ 
in the thermodynamic limit, whereas in the considered 
equally spaced model (\ref{eq-spac}) 
the crossing occurs at a value of $g$ which 
is instead $\propto N$. 

\section{Coupled Richardson Hamiltonians}\label{sec:coupled}

In this Section we introduce the model studied in the rest of the paper 
featuring two Richardson models coupled by a tunneling term 
\cite{J62,BP82}: 
\begin{equation}\label{CoupledH}
H = H_R + H_L +  H_T,
\end{equation}
where $H_R$ and $H_L$ are the ``right'' and ``left'' Richardson Hamiltonian, 
written in terms of the right and left operators $b_{\alpha,R}, b_{\alpha,L}$ 
(the fermionic operators will be denoted by $c_{\alpha \sigma,R}$ and 
$c_{\alpha \sigma,L}$ with $\sigma=\uparrow,\downarrow$). 
We consider the simpler setting in which the two models 
have the same value of the coupling $g$ and 
the same energy levels $\varepsilon_\alpha$: 
\begin{equation}\label{RichardsonHamiltonian_bosons_R}
H_s = 2 \sum_{\alpha=1}^{N} \varepsilon_\alpha b^\dagger_{\alpha,s} b_{\alpha,s}
- 2 g \sum_{\alpha,\beta=1}^{N} b^\dagger_{\alpha,s} b_{\beta,s}
\qquad\quad (s = L,\;R), 
\end{equation}
with the  $\varepsilon_\alpha$'s equally spaced and given by 
(\ref{eq-spac}). The number of levels is taken to be equal to $N$ both 
for the left and the right systems. The total number of pairs in the 
system is denoted by $M_T$ - we will also denote by 
$M_L$ and $M_R$ the operators of 
the number of pairs in the left and right system: $M_s=\sum_{\alpha=1}^{N} 
b^\dagger_{\alpha,s} b_{\alpha,s}$ (with $s = L,R$).

We write the tunneling term describing the hopping of a single fermion from one 
system to the other in the form
\begin{equation}\label{Tunneling_Hamiltonian}
H_T = - \eta 
\sum_{\sigma=\uparrow,\downarrow}\sum_{\alpha,\beta=1}^{N}
\left( c^\dagger_{\alpha\sigma,L}c_{\beta\sigma,R} +h.c.\right)
\end{equation}
(with $\eta$ having the dimension of an energy). 
Following the usual approach initially introduced by Josephson \cite{J62}, 
using second-order perturbation theory one can derive 
an effective Hamiltonian for small values of $\lambda$ (corresponding 
to the regime of weakly coupled Richardson Hamiltonians): it turns out 
the this effective Hamiltonian can be written only in terms 
of the pair operators \cite{GSD04}, greatly simplifying the study of the 
dynamics. 

Since the uncoupled Hamiltonians 
(\ref{RichardsonHamiltonian_bosons_R})
contain only interactions among pairs, 
the eigenstates of (\ref{RichardsonHamiltonian}) can be classified in terms 
of their {\em seniority} $\nu$, i.e., the number of the unpaired electrons. 
The second order effective tunneling term can be written as:
\begin{equation}\label{scdordTunn}
H^{(2)}=-\sum_\nu \sum_{\sigma=\uparrow,\downarrow}\sum_{\alpha,\beta=1}^{N} H_T 
\frac{\left|\alpha_L\beta_R\sigma;\nu\right\rangle\left\langle\alpha_L\beta_R\sigma;\nu\right|}{E_{\alpha_L\beta_R \nu}} H_T,
\end{equation}
in which the sum runs over all the possible intermediate states 
that can be reached from a $\nu$-seniority couple of states
$\left|N/2+\nu\right\rangle_L \otimes \left|N/2+\nu\right\rangle_R$, 
by removing an electron of spin $\sigma$ from the level $\beta_R$ 
of the right grain and adding it on the level $\alpha_L$ 
on the left grain (or viceversa). 
In (\ref{scdordTunn}) the quantity $E_{\alpha_L\beta_R \nu}$ is the corresponding 
excitation energy relative to the initial state.

Following \cite{GSD04}, it is possible to limit the space of states 
on which the intermediate sum runs over to the lowest energy ones, 
when acting with (\ref{scdordTunn}) on the lowest-energy states of the two 
uncoupled systems in which all fermions are bound into Cooper pairs. 
In facts, the energy $E_{\alpha_L\beta_R \nu}$ 
includes the energy needed to break 
a pair and the effect of the blocking of the states 
on the collective excitations on both subsystems.

\begin{figure}
 \begin{center}
\includegraphics[width=0.2\textwidth]{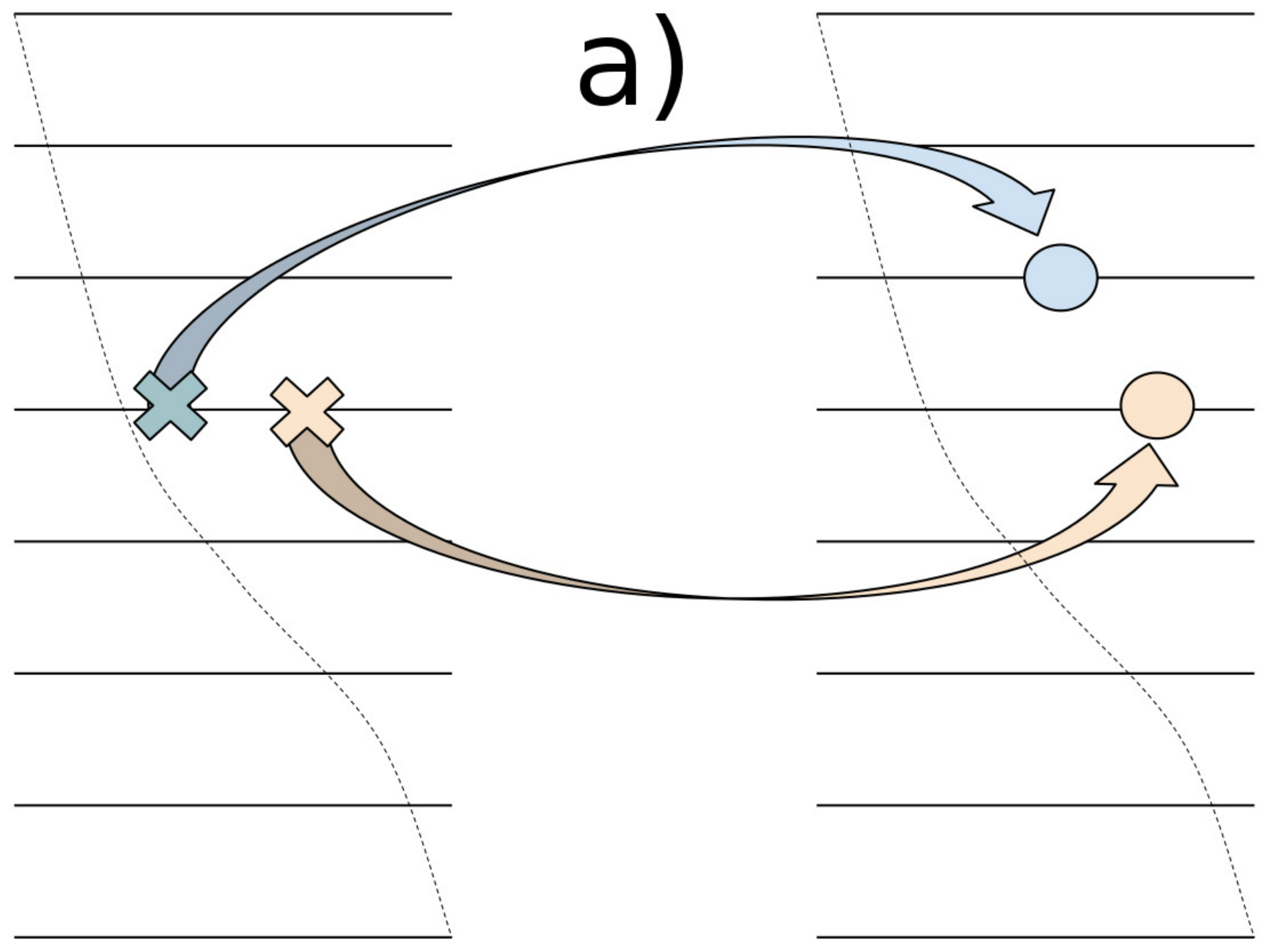}
\hspace{1.cm}
\includegraphics[width=0.2\textwidth]{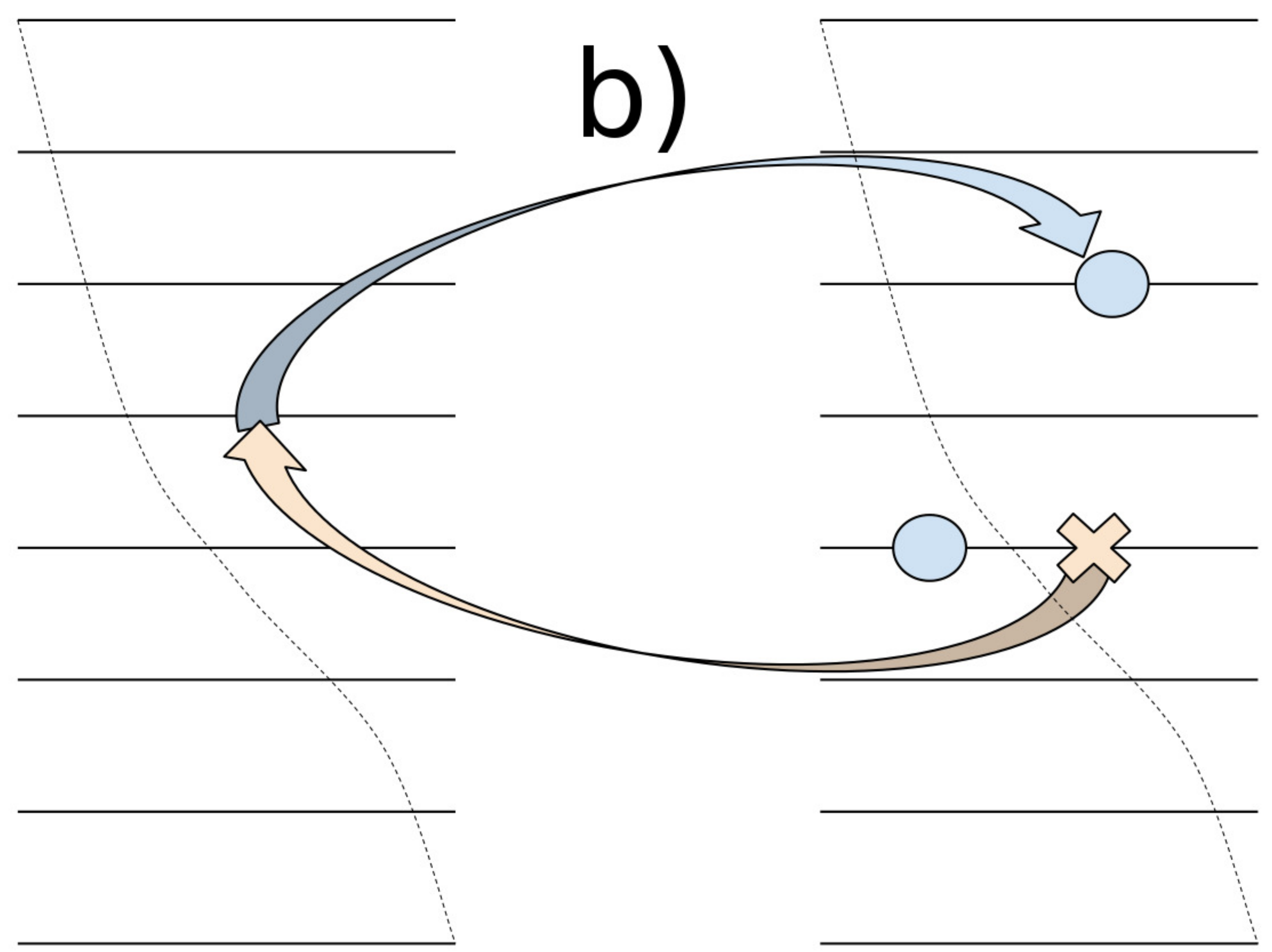}
\hspace{1.cm}
\includegraphics[width=0.2\textwidth]{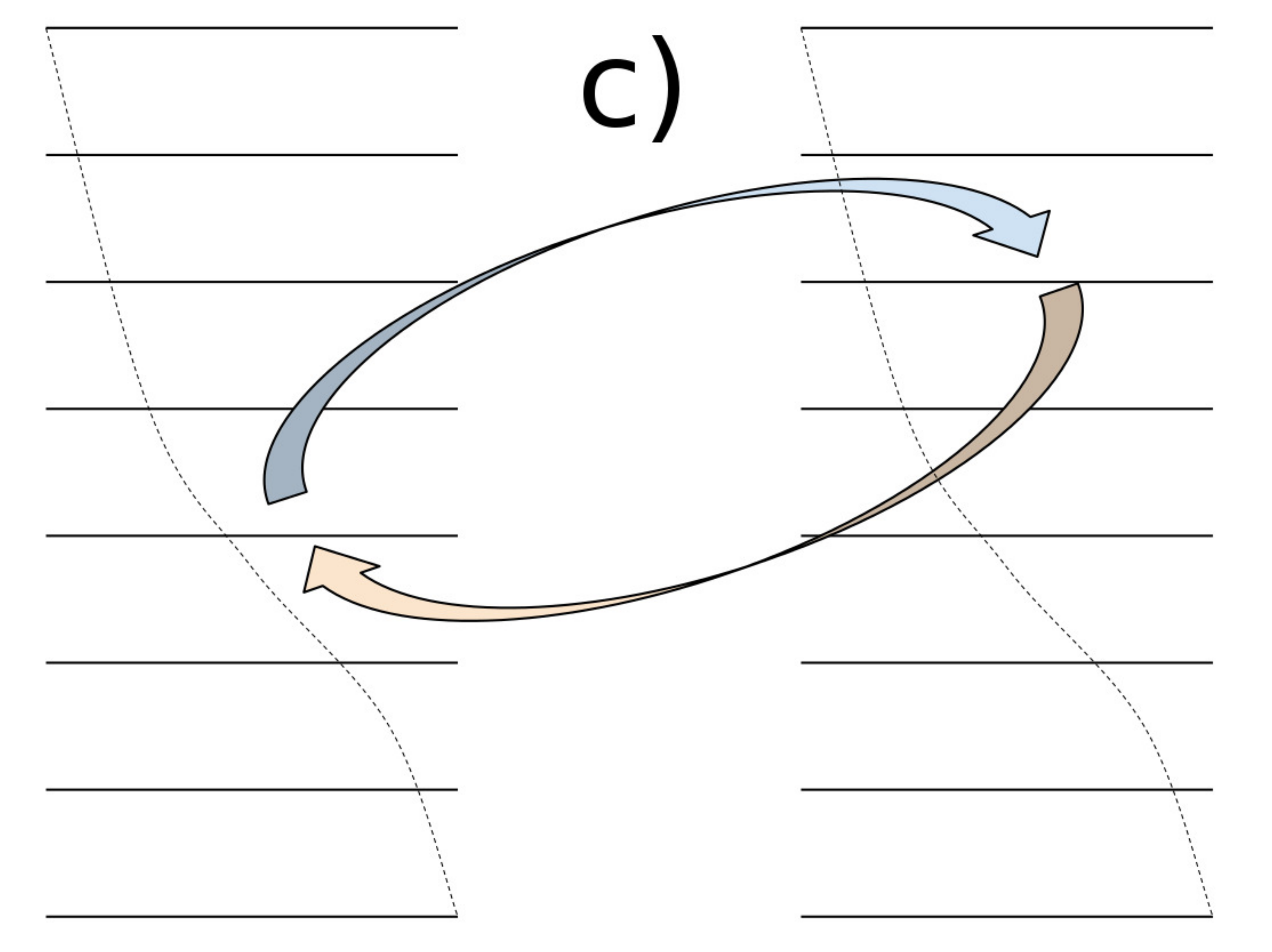}
\hspace{1.cm}
\includegraphics[width=0.2\textwidth]{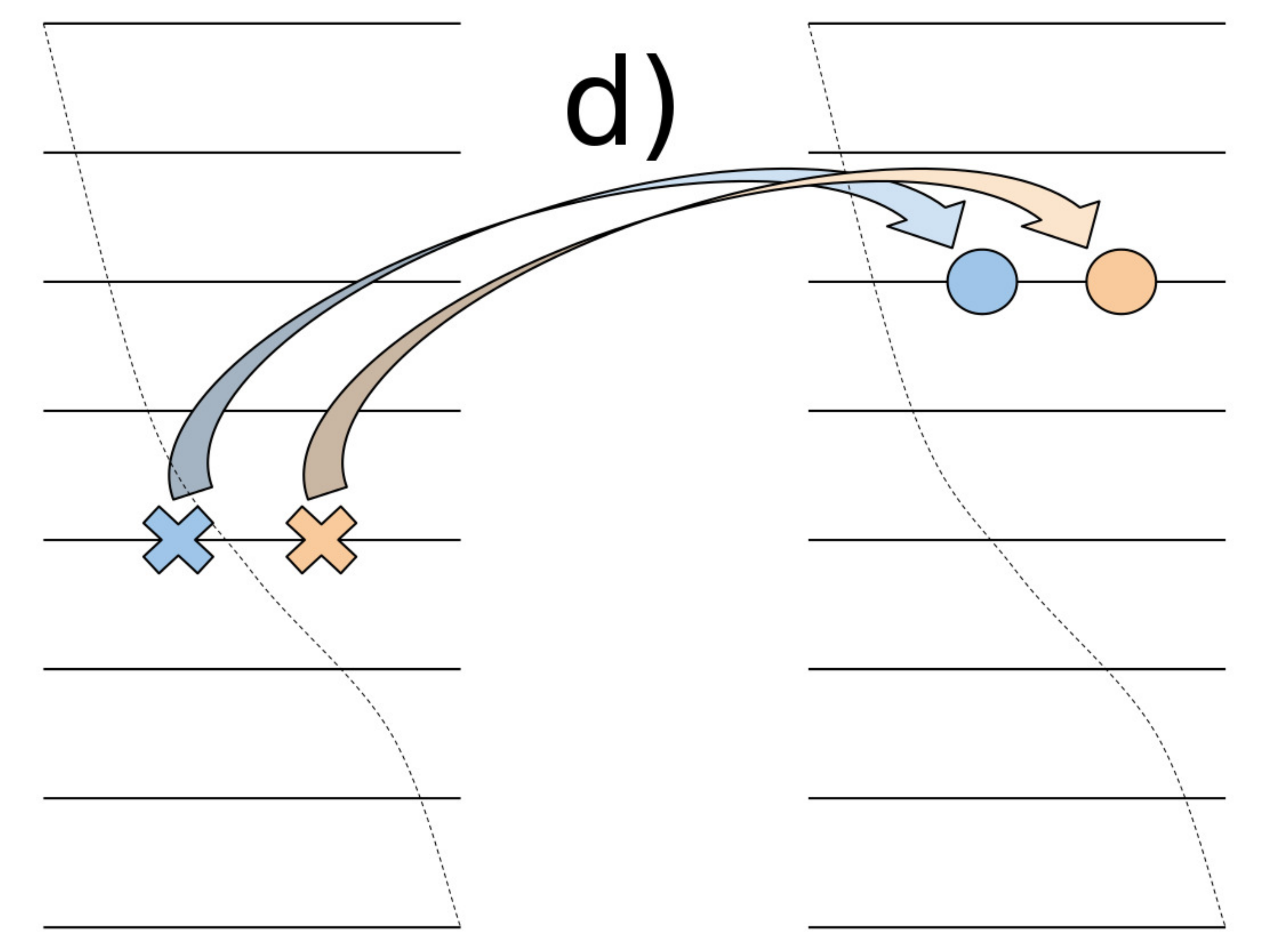}
\caption{Second-order processes associated to the Hamiltonian 
(\ref{scdordTunn}). }\label{fig:processes}
\end{center}
\end{figure}

Across the whole BCS-BEC crossover, the breaking of a pair associated 
with the tunneling of a single electron is energetically depressed: 
processes like the ones depicted in Figure \ref{fig:processes}(a)-(b)  
are suppressed by a factor $1/\DBCS$ in the dynamics, 
since they involve both the breaking of a pair, 
with an energy cost equal to the BCS gap $\DBCS$ 
and the blocking of a level, which affects all the levels 
and has therefore an energy cost roughly proportional to $N$. 
At second order in the fermion tunneling, it is more convenient 
to reach a final state in which only Cooper pairs are present \cite{J62}. 
Moreover, single-fermion tunneling does not produce a 
current in the absence of an applied driving force, so they will not 
affect the current. This is true in particular for processes like the 
one in Figure \ref{fig:processes}(c), which reproduces 
the initial state and can be included in a redefinition 
of the energies of the unperturbed system. 
We are therefore led to consider as dominant the coherent pair tunneling, 
which involves both the electrons of a Cooper pair and can 
be directly written in terms of the bosonic operators 
$b^\dagger_{\alpha,L}, b_{\beta,R}$ or $b_{\alpha,L}b^\dagger_{\beta,R}$, 
as in Figure \ref{fig:processes}(d). 
Assuming the two systems to have a definite relative phase 
(as it will be checked and discussed in Section \ref{sec:phase}), 
the coherent tunneling involves a phase shift on the state in which 
it takes place and a corresponding variation 
of the relative number of fermions $\delta N_f = \pm 2$ [see Figure \ref{fig:processes}(d)]. 

We therefore focus on coherent pair tunneling, 
for which the effective Hamiltonian is \cite{GSD04,J62} 
\begin{equation}\label{pair-tunnel}
H^{(2)} \approx -\lambda \DBCS \sum_{\alpha,\beta}
\frac{b^\dagger_{\alpha,L} b_{\beta,R}+b_{\alpha,L} b^\dagger_{\beta,R}}
{E_\alpha+E_\beta},
\end{equation}
where $E_\alpha=\sqrt{\xi_{\alpha}^2+\Delta_{BCS}^2}$, 
$\xi_{\alpha}=\varepsilon_\alpha-\mu$ and 
$\lambda = 2\eta^2/\DBCS$. 
As can be seen in Eq. (\ref{pair-tunnel}), we decided to scale 
the tunneling coefficient with $\DBCS$ since this is 
the relevant scale throughout the crossover and it is a finite quantity 
in the thermodynamic limit; furthermore, the form (\ref{pair-tunnel}) 
ensures that the tunneling acts as a perturbation also on the 
BCS side and for small values of $g$.

The form (\ref{pair-tunnel}) is particularly relevant since 
it formalizes the fact that preparing the system in its ground 
state and adding a weak fermionic tunneling term to the uncoupled 
Richardson Hamiltonians does not destroy the Cooper pairs picture: 
this provides a major simplification in the problem, allowing for 
to study the Josephson problem only in terms of hardcore bosons since 
the subspaces with different seniority will not be accessed neither 
by the ``single-site'' dynamics of the uncoupled Richardson systems, 
nor by the coupling between different sites 
(see a discussion on the single-fermion 
tunneling effects at the end of Section \ref{sec:phase}).

The ground-state state of Hamiltonian $H=H_L+H_R+H^{(2)}$ was 
studied in \cite{GSD04} and the behavior of the Josephson energy 
investigated as a function of the level spacing. Integrable versions  
of coupled pairing Hamiltonians was proposed and studied in 
\cite{ALO01,LZMG02,GFLZ02,LEDPI06}, while an analysis of the spectrum 
of two weakly coupled Richardson Hamiltonians with $H_T \propto 
\sum_{\alpha,\beta} \left( 
b^\dagger_{\alpha,L} b_{\beta,R}+b_{\alpha,L} b^\dagger_{\beta,R} \right)$ 
was presented in \cite{GSD04,KNA11}.

In the following we consider the Hamiltonian $H=H_L+H_R+H^{(2)}$, 
with $H^{(2)}$ given by (\ref{pair-tunnel}) and we investigate the properties 
of its spectrum and the dynamics starting from an state at time $t=0$ 
having an initial relative phase difference and/or an initial population 
imbalance: we are interested to ascertain for what values of $N$ a relative 
phase difference $\delta \phi$ is well defined, and to study the dynamics 
in terms of the time evolution of $\delta \phi(t)$ and $\delta M(t)$, where 
$\delta M$ is the difference between the number of pairs of the two 
systems defined by Eq.(\ref{delta_M}). 
 Note that, in general, the gap and the chemical potential appearing in
(\ref{pair-tunnel}) will be functions of time as well. However,
 for the sake of simplicity, we will consider them as constant,
 which in the present case stands as an approximation valid for
 small fractional population imbalance.

The initial state is prepared in the following way (see more details 
in Section \ref{sec:phase}): 
the uncoupled system ($\lambda=0$) is initially in the ground state, 
characterized by a definite occupation number on the left and 
on the right parts - then, at time $t=0$, the coupling $\lambda \ll d$ is 
turned on and the quantum dynamics is studied.

Integrability plays an important role both in the study of spectrum and 
dynamics: it gives the exact eigenstates of the two uncoupled systems 
and, most importantly, the exact hopping matrix elements. It also provides 
an efficient truncation mechanism to select the most important eigenstates 
in the dynamics: as we discuss in the following, one can avoid to diagonalize
the Hamiltonian written on a basis of the full 
Hilbert space of the coupled problem
and instead limit its size by restricting only to a subset of states.

Operatively, we start from the basis of the exact eigenstates 
of the two uncoupled models, with $\lambda=0$, with the number of pairs 
on the left ($M_L$) and on the right ($M_L$) separately conserved. 
Once fixed the total number $M_{T}$ of pairs, the factorized basis 
$\mathcal{S}_{M_T}$ is split into subsectors, each of them 
characterized by the occupation number of the left model $M_L$ and 
that of the right one $M_R$, such that  $M_L+M_R=M_T$. 
Denoting by $S_M$ a basis of eigenstates of 
(\ref{RichardsonHamiltonian_bosons}) 
for the subspace with given number $M$ of pairs, 
the fixed-number subspaces are spanned by
\begin{equation}\label{subbasis}
\mathcal{S}_{M_L,M_R} = \left\{ \Phi_{L}^{(M_L)} \otimes \Phi_{R}^{(M_R)} | \Phi_{L}^{(M_L)} \in S_{M_L} \,,\;\Phi_{R}^{(M_R)} \in S_{M_R} \right\},
\end{equation}
so that the factorized basis is
\begin{equation}\label{factorized-split}
\mathcal{S}_{M_T}=\bigcup_{M=\max(0,2N-M_T)}^{\min(N,M_T)} 
\mathcal{S}_{M,M_T-M}.
\end{equation}

It is possible to show that many states in $S_M$ are effectively not involved 
in the dynamics and consequently reduce the space of quantum states 
to a computationally manageable size: to see this, let first 
consider the limits $g\to 0$ and $g \to \infty$. 
In the non-interacting case, the single-level occupation numbers 
are good quantum numbers for the system: it follows that all 
the excitations above the Fermi sea ground state induced by the coupling, 
in the regime in which the tunneling coupling is small ($\lambda/d\ll 1$), 
are the states in which one particle is missing from the Fermi sea or one 
particle is added above it. 
These are a subset of the ``particle--hole'' states, 
obtained from exciting one pair from below to above the Fermi level, 
which are instead there at second order. 

In the opposite limit $g \to \infty$, it is useful to rewrite 
(\ref{RichardsonHamiltonian_bosons}) in terms of spins, obtaining 
the spin Hamiltonian (\ref{RichardsonHamiltonian_spins}): 
in the strong coupling limit $g \to \infty$, the Hamiltonian 
(\ref{RichardsonHamiltonian_spins}) reads \cite{YBA03}
\begin{equation}\label{RichardsonHamiltonian_sc}
H \approx -2g\left( \vec S\cdot \vec S -\left(S^z\right)^2-S^z \right)
\end{equation}
(where $\vec S=\sum_\alpha \vec{S}_\alpha$) and it 
conserves the total spin of the state and its $z$-projection. 
Numerical solutions of the Richardson equations show
 that the rapidities can either diverge proportionally to $g$ or remain finite, 
with real part which lies ``trapped'' between two energy levels.
 In the strong coupling limit, the tunneling Hamiltonian. Consequently, 
the states group into highly degenerate total spin 
subspaces \cite{YBA03}. In the strong coupling limit, the tunneling Hamiltonian 
(\ref{scdordTunn}) simplifies as 
well: the BCS gap diverges linearly with $g$ and all the pairs of levels 
in (\ref{pair-tunnel}) factorize a common term, yielding
\begin{equation}\label{pair-tunnel-sc}
H^{(2)} \approx - \frac{\lambda\DBCS}{\sqrt{\Delta_{BCS}^2+\mu^2}} 
S^+_{tot,L}S^-_{tot,R}+h.c.
\end{equation}
(where $\vec{S}_L=\sum_\alpha \vec{S}_{\alpha,L}$ and 
where $\vec{S}_R=\sum_\alpha \vec{S}_{\alpha,R}$). 
The ground state is the unique state in which all the rapidities 
diverge in the strong coupling limit and it is the one with highest 
(total) spin. 
The relation between the number $r$ of diverging 
roots at strong coupling and the eigenvalues of the spin 
Hamiltonian (\ref{RichardsonHamiltonian_sc}) is $r=s-s^z$ \cite{YBA03}, 
being $s(s+1)$ and $s^z$ the total spin projection along the $z$ axis. One then 
sees that it is sufficient to restrict the single-site Hilbert 
space to the root configurations with one less (or one more) 
rapidity and the same number of rapidities which diverges at 
large $g$, i.e., again the ground state of the new sector: 
therefore, no new state is needed. Although the previous arguments 
are valid in the two limiting regimes $g\to 0$ and $g \to \infty$, 
we numerically compared the results with exact diagonalization (for $N=6$) 
or the effect of adding more total spin subspaces to the dynamics (for $N=8$). 
In all the tests we performed, results in excellent agreement were found. 

\begin{figure}
\begin{center}
\includegraphics[width=0.09\textwidth]{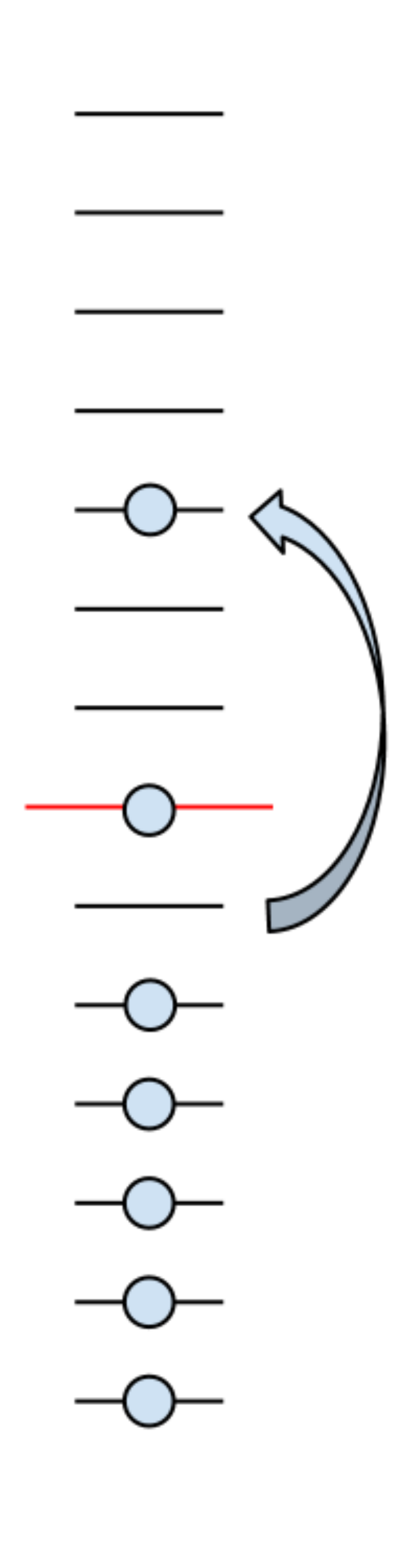}
\caption{Particle-hole states.}\label{fig:part-hole}
\end{center}
\end{figure}

Algorithms for connecting the number of roots that eventually 
diverge to the initial state configurations have been given in 
\cite{FCC08,SRD03}: 
the included states are exemplified in Figure \ref{fig:part-hole} 
and consist of the evolution in $g$ of all the configurations in which, 
in the weak coupling limit, one particle is excited from the Fermi sea 
to right above its surface or from the Fermi energy 
to one more energetic state.

The algorithm used for solving the Richardson equations numerically 
is based on the one described in \cite{BDS11}. 
To obtain eigenstates at a given $g$, one starts from $g=0$, 
where the rapidities that solve the Richardson equations are known within 
good approximation. It is therefore possible to solve numerically 
(\ref{RichardsonEquations}) for some values of $g$ around zero: 
the coefficients of the polynomial 
having these rapidities as roots are computed. 
One then extrapolates these coefficients to a 
new value of the pairing, in steps $\delta g \simeq 0.01d$,  
and compute the roots of the extrapolated polynomial, 
using them as a starting guess for the numerical solution of 
the Richardson equations. The procedure is iterated up 
to the desired value of $g$ \cite{BDS11} (see more details in \cite{B12}).
This algorithm allows to solve every configuration for sizes 
$N \le 10$, which we used in this paper. A numerical procedure for 
dealing with general Gaudin models has been presented in \cite{FEASG11}.
Once that the factorized basis has been determined, we write the 
tunneling term by using Eq. (\ref{SpSmFF-Rich}), 
and diagonalize the resulting Hamiltonian.

\subsection{Properties of the spectrum}\label{sec:spectrum}

\begin{figure}
\begin{center}
 \includegraphics[width=0.32\textwidth]{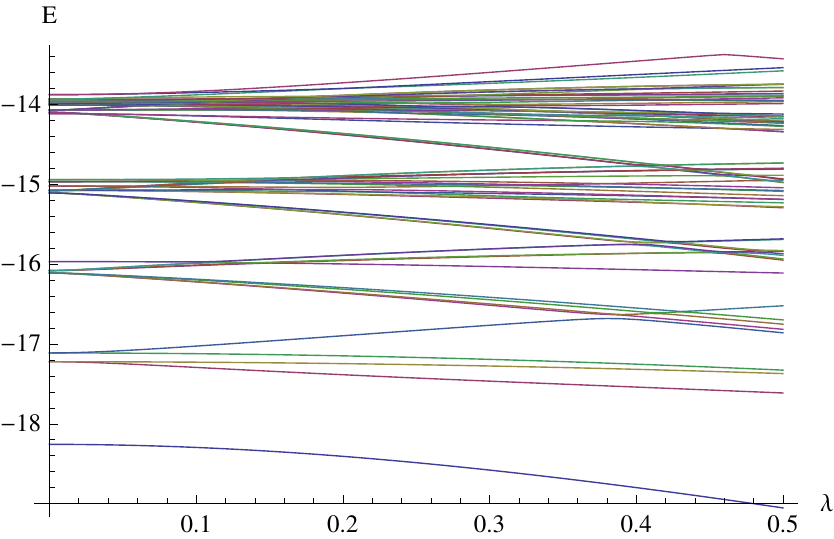}
 \includegraphics[width=0.32\textwidth]{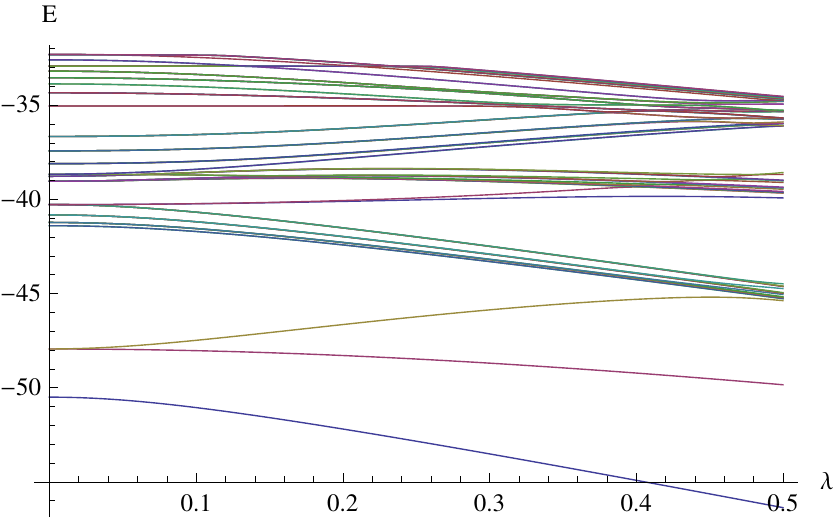}
 \includegraphics[width=0.32\textwidth]{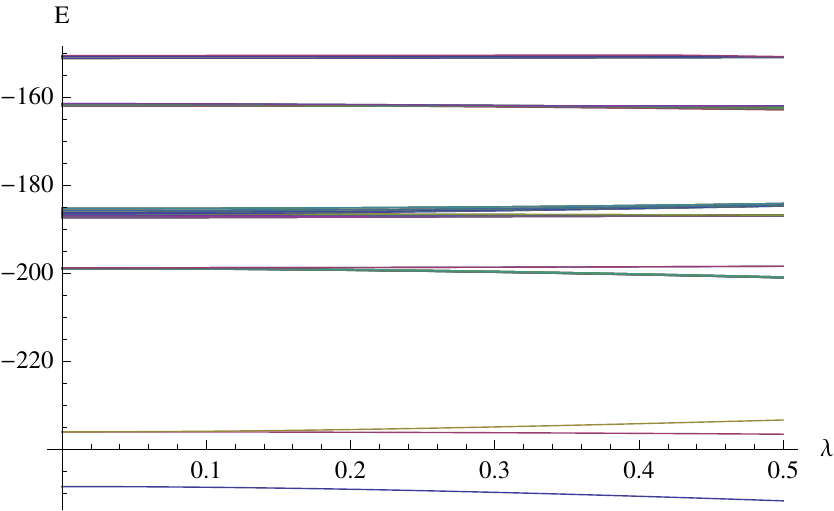}
\caption{Energy spectrum vs. $\lambda$ for $N=8$ and $M_T=8$, 
with $g=0.1$ (left), $g=1.2$ (center), $g=6.2$ (right) - for simplicity, 
thereafter also $\lambda$ is expressed in units of $d$.}
\label{fig:spectrum}
\end{center}
\end{figure}

In Figure \ref{fig:spectrum} we plot the energy 
spectrum for two coupled Richardson models as a function of the 
tunneling parameter $\lambda$ for three different values of $g$. It is seen 
that the effect of a weak tunneling on the spectrum 
depends essentially on the coupling among fermions: 
one can clearly identify a regime of nearly non-interacting particles, 
in which the quasi-degeneracy of the levels is given by the number of ways 
of promoting one or more particles in an excited level to obtain 
a given energy (degeneracy is a consequence of the choice of equally-spaced levels). 
In this regime, the perturbation splits the levels of one band 
as far as the band spacing, hence giving rise to a spectrum 
in which the original degeneracies are not seen any more.

On the other hand, in the strong coupling regime states 
group into eigenstates of the total angular momentum, as
seen from the spin representation (\ref{RichardsonHamiltonian_sc}). 
Since the distance among the energies of these subspaces is of order $g$, 
in this regime, even a tunneling term of several times the gap 
cannot mix the different subspaces among them. 

In the crossover region, the strong coupling subspaces 
are already quite defined, but not far one from the other. 
It follows that a sufficiently strong perturbation can still hybridize them. 
To better illustrate this point, we may evaluate how much the levels 
are shifted by turning on $\lambda$: however, the absolute value of 
the shift should be compared with the level spacing in a situation 
where levels are well-distinguishable (intermediate couplings) and 
the band spacing in the presence of strong degeneration 
($g\to0$ or $g\to\infty$). 

\begin{figure}
 \begin{center}
\includegraphics[height=0.2\textheight]{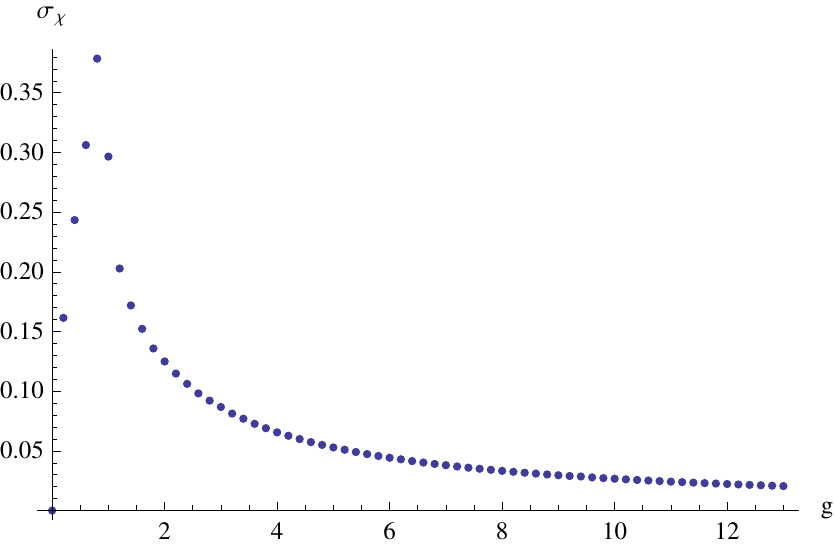}
 \includegraphics[height=0.2\textheight]{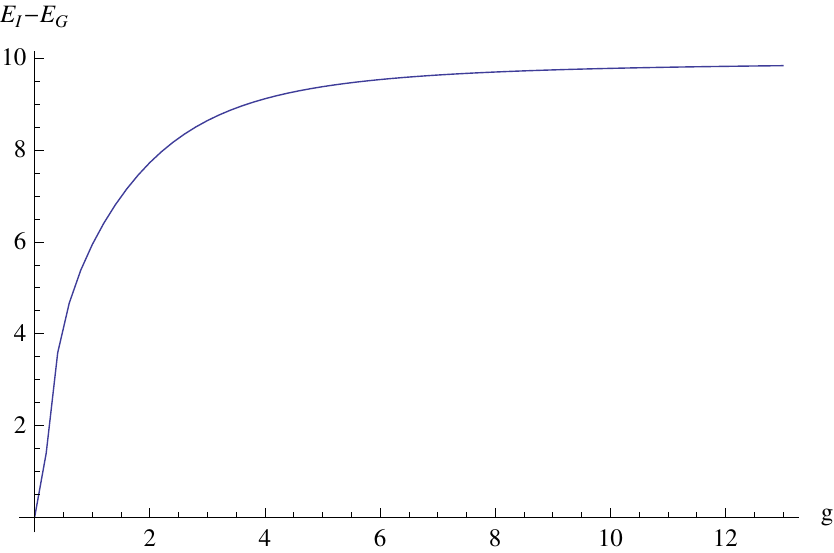}
\caption{Left: standard deviation $\sigma_\chi$ vs. $g$ 
associated with the level structure, given by 
($N_{bin}=100$, $\lambda=0.05$, $N=8$, $M_T=6$). 
Right: energy difference between the first excited state and 
the ground state as a function of $g$ for the same values of the parameters.}
\label{fig:susc}
\end{center}
\end{figure}

A convenient way to proceed is to divide the all spectrum in a certain number 
of interval (let $N_{bin}$ this number) and count how many levels lie in each 
interval: we define the quantity
\begin{equation}
\chi_m \equiv (\#\;levels\; in \; the \; m^{th} \; interval)(\lambda)-
(\#\;levels\; in \; the \; m^{th} \; interval)(\lambda=0).
\label{chi}
\end{equation}
The average of this quantity 
with respect to the interval index $m$ is obviously zero. 
The relevant quantity is instead its 
 standard deviation $\sigma_\chi$: the result is plotted 
in Figure \ref{fig:susc}, left part. 
The quantity $\sigma_\chi$ has a maximum around $ g/d\sim 1$, 
corresponding to the point in which the degeneracies of the 
noninteracting picture are already destroyed, 
while the energy bands of the strong coupling regime are not evident yet. 
This result is of course related to the occurrence of a crossover 
between weakly attracting fermions and tightly bound pairs: 
for the uncoupled systems, from the point of view of the energy spectrum
the crossover reflects itself on the creation of energy bands out of 
the pair levels, which are more and more separated by increasing $g$. 
This is also seen at the level of the coupled spectrum, 
The doublet structure characterizing coupled noninteracting systems 
is melted into an highly-degenerate band structure.

In the right part of Figure \ref{fig:susc}, 
we plot the energy difference between the first excited state and 
the ground state in a system with an odd number of particles as a function of $g$. 
It turns out that, as long as the gap opens more and more, 
the energy difference between the components of the level doublet 
reaches a maximum splitting.

\section{Emergence of a definite relative phase}\label{sec:phase}

In this Section we explain what initial states have been considered and we 
discuss how a definite relative phase emerges for small number of 
particles and our algorithm for obtaining it from numerical data. 

The initial state $\left|\Psi(t=0) \right\rangle$ is prepared in 
a linear combination of ground states of the 
uncoupled systems having a different number of pairs 
and therefore a population imbalance: this state is evolved in time 
with the dynamics generated by (\ref{CoupledH}), 
with $\lambda \neq 0$ in the tunneling term. 
More precisely, at $t=0$ the system is generically in a linear superposition 
of two states with a given number of pairs (we choose $M_0$ in a system 
and $M_0-D$ in the other): the total number of pairs is conserved during the 
time evolution and is $M_{T}=2M_0-D$. 
Denoting by $\left| \Phi_{M}^{(L,R)} \right\rangle$ 
the lowest-energy state with $M$ pairs of either the left or the right system, 
we prepare the system in the state
\begin{equation}\label{initialstate}
\left| \Psi(t=0)  \right\rangle = \frac{1}{\sqrt{1+\xi^2}}\left(
 \left| \Phi_{M_0}^{(L)} \right\rangle \otimes \left| \Phi_{M_0-D}^{(R)} \right\rangle 
 + e^{i\phi_0}\xi\left| \Phi_{M_0-D}^{(L)} \right\rangle \otimes \left|  
\Phi_{M_0}^{(R)} \right\rangle 
\right):
\end{equation}
given the limitation on the total number of pairs 
($M_T\le 10$), we will most consider $D=1$, therefore creating 
as initial state by a linear superposition of $M_0$ in a system 
and $M_0-1$ in the other, and $D=2$ (typically we choose $M_0=N/2$ or $M_0=N/2\pm 1$).
 
Given $\left|\Psi(t) \right\rangle$, one can compute the pair 
population imbalance as 
\begin{equation}\label{delta_M}
\delta M(t)  \equiv \left\langle \Psi(t) \left| M_L-M_R\right| \Psi(t) 
\right\rangle = \left\langle \Psi(t) 
\left| \sum_\alpha \left( b^\dagger_{\alpha,L} 
b_{\alpha,L} - b^\dagger_{\alpha,R} 
b_{\alpha,R} \right) \right|  
\Psi(t) \right\rangle
\end{equation}
(the population imbalance $\delta N_f$ is just $\delta N_f=2 \delta M$). Notice that with $\xi=0$ it is  
$\left\langle \Psi(t=0) \left| M_L \right| \Psi(t=0) 
\right\rangle=M_0$.

According to the notation of \cite{SFGS97}, 
we will denote by $z(t)$ the fractional population imbalance: 
\begin{equation}
z(t)=\frac{\delta M(t)}{M_T}.
\label{def_z}
\end{equation}
Using the state (\ref{initialstate}) one simply obtains
$$\delta M(t=0)=D \, \frac{1-\xi^2}{1+\xi^2}:$$ 
therefore varying the parameter $\xi$ one can choose different initial 
population imbalances (with $\left| \delta M(t=0) \right| \le D)$. 
The dynamics is then studied turning 
on a small perturbation ($\lambda / d = 0.01 - 0.1$ in our runs) 
and compute the time evolution of the state after exact diagonalization 
the Hamiltonian. The main limitation of this protocol arises 
from the consume of RAM by diagonalization subroutines: 
by limiting subspaces appropriately, 
as discussed in Section \ref{sec:coupled}, 
one can study systems up to $N=10$ levels (both on left and 
right systems).

An important issue we want to address in this Section,
 arising from the fact that we can treat the exact quantum 
 dynamics of the coupled model only for a limited number of pairs,
 is whether a definite relative phase emerges at small sizes.
 In the presence of the tunneling term (\ref{scdordTunn}), 
eigenstates will in principle be written as a combination of 
many of the factorized states of the two uncoupled Hamiltonian. 
Nevertheless, as we will see, when the initial population imbalance is 
small, the number of involved states is rather small. 
Moreover, even for higher particle imbalance, when the tunneling 
is weak and the pairing strength is strong enough, 
the Hilbert space of each system organizes in subspaces, labeled 
by eigenvalues of the total spin (see Section \ref{sec:Richardson}). 
It follows that in most cases, even if the exact states involved 
are many, the corresponding energy eigenvalues are not very different, 
 therefore the time evolution takes place with nearly definite phase.

Note that in our canonical setting the expectation value 
$\left\langle \Psi(t) \right| b_{\alpha,L/R} \left| \Psi(t) \right\rangle$ 
is always vanishing, since the $b$'s operators does not conserve the 
number of particles. However, we can define time-dependent 
phase differences 
between one level in a system and a level in the other system systems  
by the use of the formalism of Section \ref{sec:Richardson} evaluating 
the dynamical two-point functions (for the uncoupled systems, 
dynamical two-point correlations have been studied \cite{ZLGM03,FCC10}). 

\begin{figure}
\begin{center}
 \includegraphics[width=0.45\textwidth]{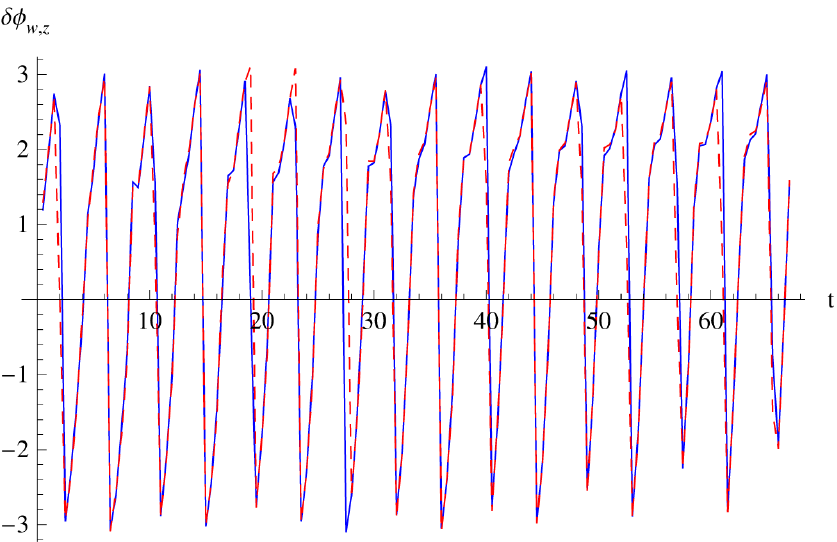}
\end{center}
\caption{Phase differences $\delta \phi_{w}(t)$ (solid blue line) 
and $\delta \phi_{z}(t)$ (dotted red line)  
vs. $t$ for the coupled systems with $N=8$ levels each, 
total number of pairs $M_T=8$, pairing strength $g=0.6$, 
tunneling parameter $\lambda=0.1$, $D=2$ (corresponding to $M_0=5$) and initial imbalance 
$z(t=0)=0.25$. Time here and in the following figures is in units 
of $\hbar/d$.}\label{fig:LR&LF}
\end{figure}

From the correlation function 
$\bra{\Psi(t)} b^\dagger_{\alpha,L} b_{\beta,R} \ket{\Psi(t)}$ one can 
extract how much the phases of two distinct levels differ at a given time. 
In particular, we considered two different procedures for the choice 
of the levels, which can be tested one against the other, and define: 
\begin{equation}\label{zw_w}
w_\alpha(t)=\bra{\Psi(t)} b^\dagger_{\alpha,L} b_{\alpha,R} \ket{\Psi(t)}
\end{equation}
and 
\begin{equation}\label{zw_z}
z_\alpha(t) \equiv \bra{\Psi(t)} b^\dagger_{\alpha,L} b_{N/2,R} 
\ket{\Psi(t)}.
\end{equation}
In (\ref{zw_z}) the subscript refers to the level on the left system and a 
reference state is taken on the right system (arbitrarily chosen to be the level 
$N/2$); conversely, in (\ref{zw_w}), 
the level is chosen to be the same on both systems. We define a relative 
phase between levels as
\begin{equation}\label{zw_w_polar}
w_\alpha(t) \equiv \left| w_\alpha(t) \right| e^{i \delta \phi_{w}(t;\alpha)} 
\end{equation}
and 
\begin{equation}\label{zw_z_polar}
z_\alpha(t) \equiv \left| z_\alpha(t) \right| e^{i \delta \phi_{z}(t;\alpha)}.
\end{equation}
The functions $\delta \phi_{w}(t;\alpha)$ and $\delta \phi_{z}(t;\alpha)$ 
are functions of both time and level index. 
It is then necessary to verify whether the levels have small phase difference: 
to do this, we define the level average
\begin{equation}
\delta \phi_{w,z}(t)=\frac{1}{N} \sum_{\alpha=1}^{N} 
\delta \phi_{w,z}(t;\alpha)
\end{equation}
and their standard deviation $\sigma_{w,z}(t)$ [with 
$\sigma_{w,z}^2(t)=(1/N) \sum_{\alpha=1}^{N} 
(\delta \phi_{w,z}(t;\alpha)-\delta \phi_{w,z}(t))^2$].
The time evolution of the mean values $\delta \phi_{w,z}(t)$ is reported in Figure  
\ref{fig:LR&LF}: one sees that already for $M_T=8$, one has
relatively small values of $g$ where  the two definitions of the relative phase 
are in good agreement for most of the times. The two definitions 
$\delta \phi_{w}(t)$ and $\delta \phi_{z}(t)$ are expected to agree 
only when the two systems show coherent behavior, and 
the phase difference between them is, within a good approximation, 
given by the phase difference between any two levels chosen. We checked 
that choices other
 than (\ref{zw_w})-(\ref{zw_z}) give practically the same results when 
a relative phase is well defined.

\begin{figure}
 \begin{center}
 \includegraphics[width=0.48\textwidth]{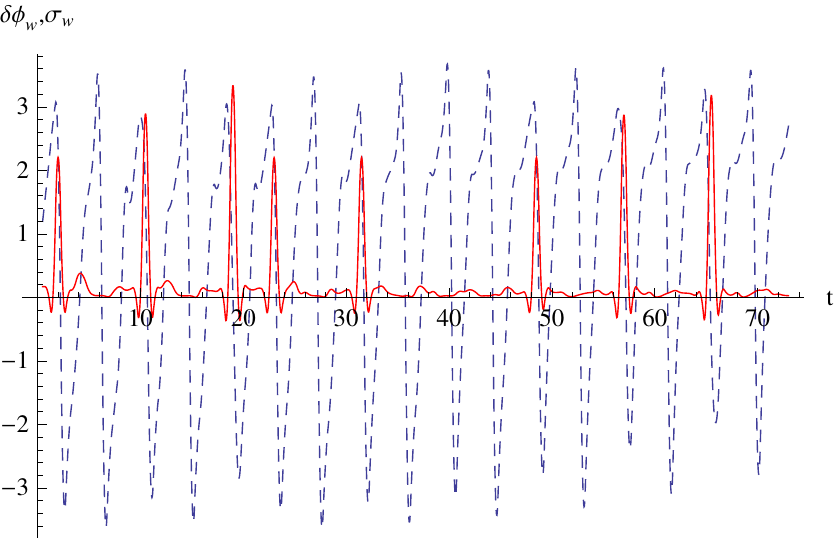}
 \includegraphics[width=0.48\textwidth]{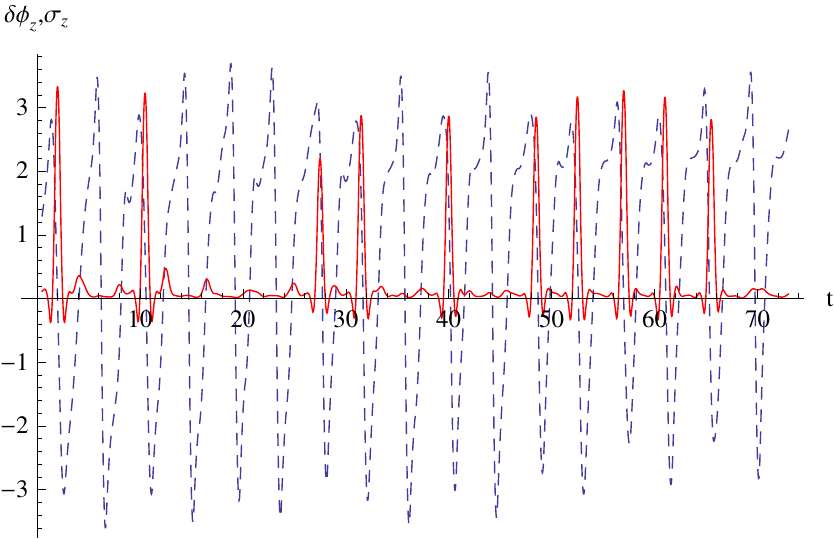}
\end{center}
\caption{Phase difference means $\delta \phi_{w,z}(t)$ 
(dashed blue lines) and standard deviations $\sigma_{w,z}$ (red solid lines), 
as determined from the correlation functions $w$ (left) and $z$ (right), 
for two coupled grains with the same parameters as in the previous Figure.
}
\label{fig:M&SD06}
\end{figure}

\begin{figure}
 \begin{center}
 \includegraphics[width=0.48\textwidth]{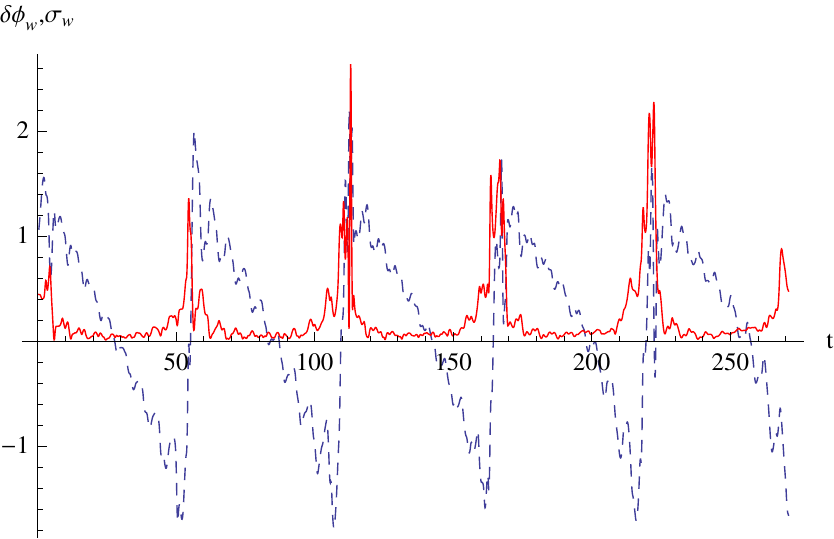}
 \includegraphics[width=0.48\textwidth]{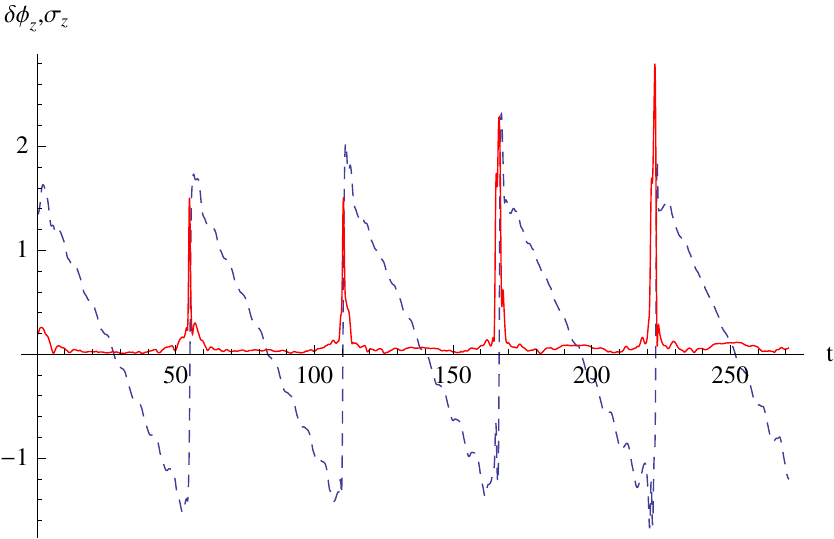}
\end{center}
\caption{Same quantities as in Figure \ref{fig:M&SD06} with $g=0.2$ 
(other parameters unchanged).}\label{fig:M&SD02}
\end{figure}

In order to have a definite relative phase one has to check that 
the average values $\delta \phi_{w,z}(t)$ 
should be (possibly for most of the considered 
times) much larger than their standard deviations $\sigma_{w,z}(t)$: as shown 
in Figures \ref{fig:M&SD06} and \ref{fig:M&SD02} 
(done respectively for $g=0.2d$ 
and $g=0.6d$) this condition is rather well verified also for a number 
of pairs $M_T=8$. One also sees that for $g=0.2d$ the agreement is less 
good, as expected also from the fact that - as discussed in Section 
\ref{sec:Richardson} - the uncoupled systems have significant deviations 
from the large-$N$ limit. We also observed for the considered values of $g$ 
a significant degradation of the relative phase for even 
smaller total number of pairs, 
e.g. as low as $M_T=4$.

Information about the phase difference averages and their standard 
deviations at every given time is useful, but we can complement it with their 
averages in time: to this purpose, we consider the mean of the 
standard deviation presented above over sufficiently long times 
(several periods) 
$$C_{w,z}^{\delta \phi}=\frac{1}{t_{max}}
\int_0^{t_{max}}\sigma_{w,z}(t')dt' .$$ 
To establish a comparison, we need to evaluate also 
the mean phase difference among the condensates. 
This is an oscillating quantity, having vanishing average on time: 
we then compute the average of its square: 
$$S_{z,w}^\phi=\sqrt{\frac{1}{t_{max}}\int_0^{t_{max}} \delta\phi_{z,w}^2(t')dt'}.$$

\begin{figure}
 \begin{center}
 \includegraphics[width=0.5\textwidth]{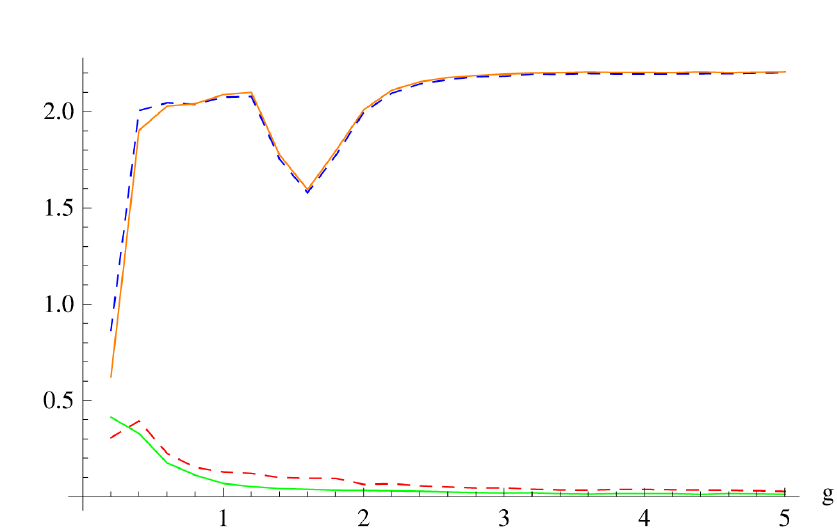}
\caption{$C^{\delta \phi}_w$ (lower dashed line), 
$S^{\delta \phi}_w$ (upper dashed line) and $C^{\delta \phi}_z$ 
(lower solid line), $S^{\delta \phi}_z$ (upper solid line) vs. 
$g$. Parameters are as in Figures \ref{fig:LR&LF}-\ref{fig:M&SD02}: $N=8$, $M_T=8$, $\lambda=0.1$,  
$D=2$ (moreover, $t_{max}=1000$).}
\label{fig:graincoherence}
\end{center}
\end{figure}

In Figure \ref{fig:graincoherence} we plot 
$C^{\delta \phi}$ and $S^{\delta \phi}$ 
with both the definitions (\ref{zw_w})-(\ref{zw_z}). 
One sees already for $g \gtrsim 0.3d$ a very good agreement 
$C^{\delta \phi}_w$ and $C^{\delta \phi}_z$, and both significantly larger 
than the time averages $S^{\delta \phi}_{w,z}$. For this reason we are going 
to denote as $\delta \phi$ the relative phase difference, omitting the 
indexes $w,z$. One also sees that for small $g$ the relative phase is not 
defined, as expected, since the relative phase is comparable with its variance.

As a function of the pairing parameter $g$, from 
Figure \ref{fig:graincoherence} one sees that 
the higher is the value of $g$, the more the system shows a
 definite relative phase. We also observed that the smaller is 
the tunneling parameter, the sooner (in $g$) a definite phase 
is established. Similarly, a small initial imbalance 
allows for a definite phase to emerge for relatively small values of $g$, 
while - for the considered values of $N$ - 
stronger pairing is necessary if states with 
larger initial imbalances are selected. 
This is due to the fact that the initial state 
is projected on few states in the lowest part of the 
spectrum when the initial population difference is small. 
Conversely, larger population imbalances at $t=0$ 
are projected to many states in the middle of the spectrum, 
each having its own energy.

We pause here to comment about fermion tunneling: as a matter of fact, 
the original tunneling 
Hamiltonian (\ref{Tunneling_Hamiltonian}) is written in terms of 
fermionic operators, while the result 
that the phase coherent behavior is established 
with relatively small pairing and/or total number of pairs 
is obtained with the bosonic approximation (\ref{pair-tunnel}), 
acting on the restricted subspace of blocked levels. 
Since at small $g$ pair-breaking excitations 
may play an important role, a natural question to ask is whether the presence 
of fermionic degrees of freedom, aside of bosonic pairs, may spoil 
the phase-coherent behavior of the systems for sufficiently large pairing. 
The issue can be rephrased into the question of whether the initial state, 
during the evolution generated by the coupled Hamiltonian, 
containing a fermionic tunneling term, may give rise to a 
huge number of states in which two or more electrons are 
not paired, evolving incoherently with respect to the states 
in which only pairs appear.

These states have to be written as linear combinations of the factorized 
states of the two uncoupled Hamiltonians. On each site, 
the energy of such states can be exactly computed for any value of $g$. 
In order to have an estimation of a lowest bound for the energy, 
we can consider a state in which the most energetic pair is broken and 
one electron is promoted into the next level, which 
reduces the number of pairs by $1$ and the number of unblocked levels 
by $2$, as seen in Section \ref{sec:Richardson}. 
The energy of the lowest pair-breaking excitation has 
been considered in \cite{YBA03} and it reads:
\begin{equation}\label{pairbreak}
E_{pair} \simeq \frac{\varepsilon_M + 
\varepsilon_{M+1}}{2} -g (M-1) ((N-2)-(M-1)+1)
\end{equation}
The bare energies in the first term of (\ref{pairbreak}) do not depend on $g$, 
unlike the ground state energy, all the pair-conserving excitations and 
the second term in the previous equation. It follows that, 
by taking the pairing strength sufficiently high, 
all pair-breaking excitations can be made to lay at arbitrary energy 
above the ground state and are therefore suppressed 
with respect to pair-conserving excitations.

Checking explicitly that the insertion of states with unpaired 
electrons does not spoil the phase relation requires 
much larger computational effort, in that the Hilbert 
space should be enlarged to the $\binom{N}{m}\binom{N-m}{M}$ 
configurations in which the $m$ electrons can ``block'' part of 
the $N$ levels, with fixed number $M$ of pairs. 
We can therefore qualitatively rely on the standard argument 
based on the presence of a gap 
preventing single-fermion tunneling: 
note that this should already hold for values of $g \gtrsim 0.25$, 
as previously discussed.

\begin{figure}
 \begin{center}
 \includegraphics[width=0.49\textwidth]{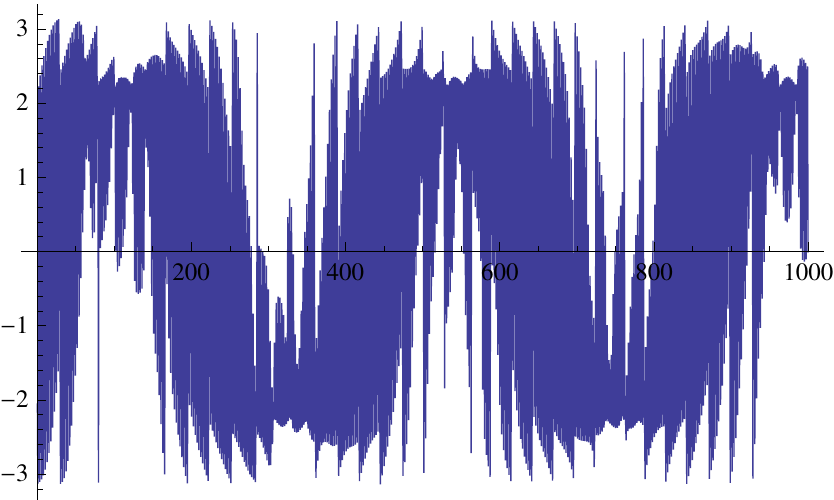}
 \includegraphics[width=0.49\textwidth]{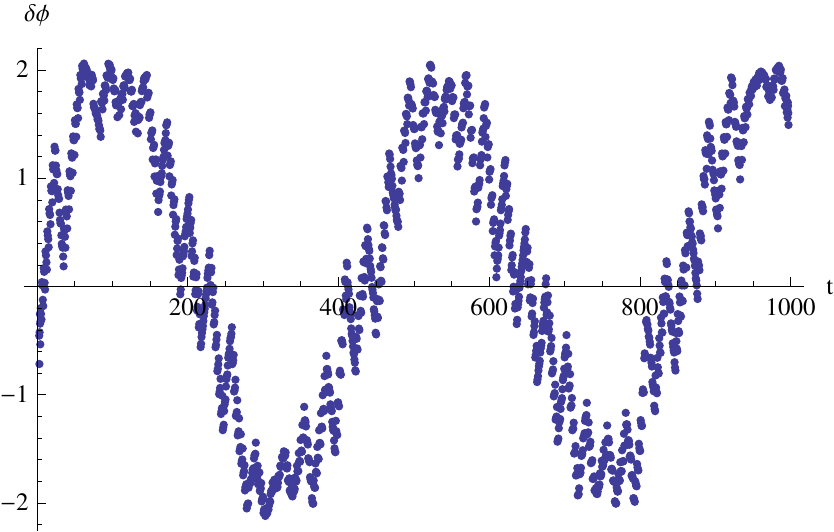}
 \caption{Left: phase difference, as a function of time, 
for $N=8$, $M_T=6$, $\delta M_0=2$, $g=5$, $\lambda=0.05$. 
Right: averaging over short times $t_{av}$ 
(here, $t_{av}=6$) 
to remove fluctuations.}\label{fig:phase}
\end{center}
\end{figure}

We also mention that, even if the phase is quite well defined, 
residual fluctuations can still be observed, in such a way that 
the widest, slowest oscillations are superimposed with faster 
and narrower ones. We find convenient to isolate the former 
ones by computing time 
averages on intervals much smaller than the period of 
the largest oscillations: this allows to better understand 
the structure of the dynamical diagrams discussed in the next Section. 
An example of the procedure is provided in Figure \ref{fig:phase}.

\section{Phase portrait and current-phase characteristics}
\label{sec:dyn}

In this Section we first draw the population-phase 
dynamical portrait $z(t)$-$\delta \phi(t)$ as it has been done 
for bosonic Josephson junctions \cite{SFGS97,RSFS99} and we determine 
the current-phase characteristics, which is a typical tool  
used to characterize the behavior of a Josephson junction \cite{L79,BP82}. 
From the solution for the quantum dynamics one can extract 
the dominant period of the population oscillations and determine 
the Josephson frequency. We also comment on the determination 
of a two-state model giving a good description of the dynamics and of 
the current-phase characteristics for the considered initial conditions. 
We observe that most of our simulations are done for the initial state 
(\ref{initialstate}) with $D=1$ build by a linear combination of a state with 
$M_0=N/2$ pairs on a system and $M_0=N/2-1$: this state has a maximum value 
for $\left| \delta M(t=0) \right|$ equal to $1$. For this initial state 
the relative phase difference $\delta \phi$ is well defined 
for a total number of pairs $\gtrsim 6$ and for $g \gtrsim 0.3$ 
(see Figure \ref{fig:graincoherence}), where the expectation values 
for the correlation functions are already 
rather similar to the large-$N$ BCS findings \cite{FCC08}: we can then explore 
the crossover region (which is around $g/d \sim 0.25 N$). In the final 
part of the Section we consider $D=2$ and initial imbalance $\delta M(t=0)=2$: 
the phase turns yet to be again rather well defined (but a larger values 
of $g$), but we cannot practically explore larger initial imbalances 
(i.e., larger values of $D$) since 
with our maximum value of pairs $M_T \sim 10$ the relative 
phase is well defined only 
for very large values of $g$ (well beyond the crossover point).

\begin{figure}
\begin{center}
\includegraphics[width=0.49\textwidth]{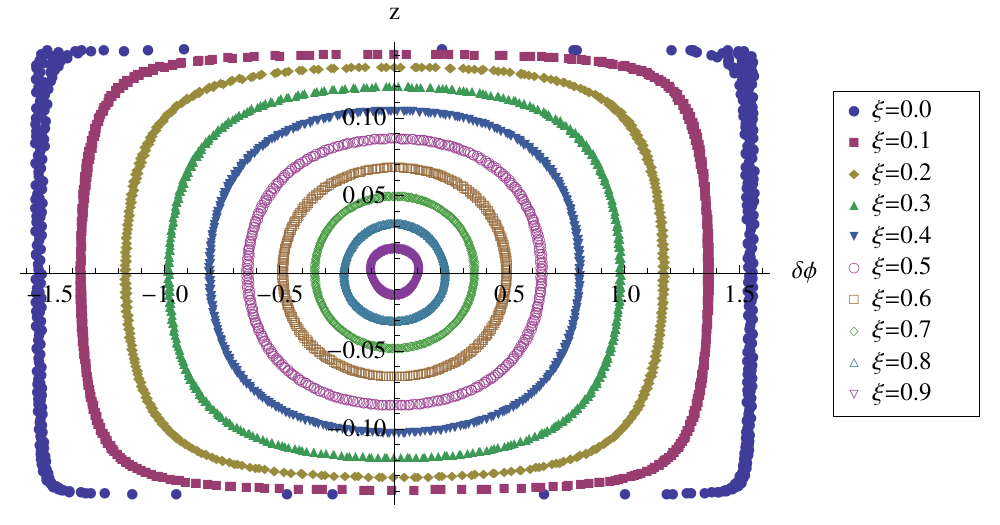}
\caption{Dynamical phase-portrait $z$-$\delta \phi$ 
for different values of the parameter $\xi$ in the BCS regime, 
with $N=8$, $M_{T}=7$, $D=1$, $g=0.57$, $\lambda=0.05$. 
The chosen values of $\xi$ are 
$\xi=0.1,0.2,0.3,0.4,0.5,0.6,0.7,0.8,0.9$ corresponding respectively 
to $\delta M(t=0)=0.98,0.92,0.83,0.72,0.60,0.47,0.34,0.22,0.10$.}
\label{fig:phasediagram-xi}
\end{center}
\end{figure}

\begin{figure}
\begin{center}
\includegraphics[width=0.55\textwidth]{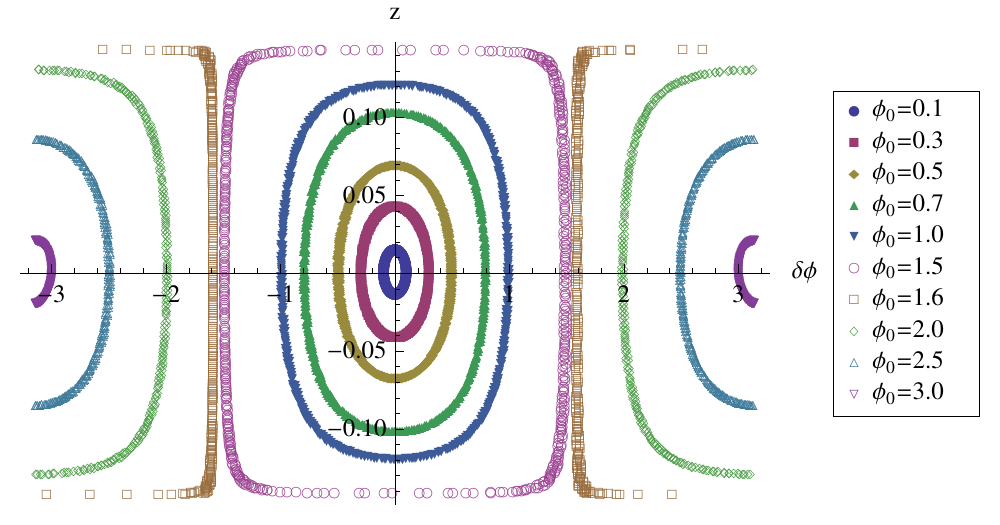}
\caption{Phase portrait for different initial phases in the BEC regime 
($N=8$, $M_{T}=7$, $D=1$, $\delta M(t=0)=1$, $g=9.7$, $\lambda=0.05$).}
\label{fig:phasedigram-phi}
\end{center}
\end{figure}

In Figure \ref{fig:phasediagram-xi} we plot the number-phase portrait 
where we plot as a function of time both $\delta \phi(t)$ and 
$\delta M(t)$ for different values of $\delta M(t=0)$. 
It is also possible to study the diagram while varying 
the initial phase in the initial state (\ref{initialstate}), as shown 
in Figure \ref{fig:phasedigram-phi}.

One sees from Figures \ref{fig:phasediagram-xi}-\ref{fig:phasedigram-phi} 
that even for a small total number of pairs ($M_T=7$) the phase diagram 
in the plane $\delta \phi-z$ shows a remarkable agreement 
with a "pendulum" law of motion in the small oscillations regime 
when the initial imbalance is small. Furthermore, 
as the initial displacement or phase difference becomes larger, 
significant corrections are seen.

A way to understand such results is to introduce a two-state model 
\cite{BP82,FLS65}: computing the overlaps of the initial state 
(\ref{initialstate}) having 
$D=1$ and $|\delta M(t=0)| \le 1$ with the 
many-body eigenfunctions of the full Hamiltonian (with $\lambda$ small), 
one sees that the largest overlaps are with the ground and the first 
excited states. 
Given this one expects that the dynamics is well explained by a simple 
linear two-mode model involving such two states. The dynamical equations of the 
Feynman two-state model are reviewed in Appendix \ref{appendix:dyn}: 
for the linear two-state model here considered the phase difference 
$\delta \phi$ does not overcome the value $\pi/2$, i.e., if 
$\left| \phi(t=0) \right|<\pi/2$, then $\left| \phi(t) \right|<\pi/2$. 
As seen in Figures 
\ref{fig:phasediagram-xi}-\ref{fig:phasedigram-phi}, this property 
is clearly observed in the numerical results (we also checked it 
with exact diagonalization). The property is typical of the linear 
two-mode model and it is connected with the fact that the main 
contributions to the time-dependent wavefunction arise from the first 
two lowest-lying states of the interacting system with equal weights. 

We now focus on the pair current between the models: we define 
the current $I$ as the 
time derivative of the occupation number of the left subsystem
\begin{equation}\label{current-def-b}
I(t) \equiv \frac{d}{dt} \left\langle \Psi(t) \left| M_L\right| \Psi(t) 
\right\rangle = \frac{d}{dt} \left\langle \Psi(t) 
\left| \sum_\alpha b^\dagger_{\alpha,L} 
b_{\alpha,L} \right|  
\Psi(t) \right\rangle.
\end{equation}
From Eq. (\ref{current-def-b}) one finds
\begin{equation}\label{current-def}
\hbar I(t) = i\left[H,M_L\right] = i\left[H^{(2)},M_L\right]=
i\sum_{\alpha,\beta}\frac{b_{\alpha,L} b_{\beta,R}^\dagger - b_{\alpha,L}^\dagger b_{\beta,R}}
{E_\alpha+E_\beta}.
\end{equation}

\begin{figure}
\begin{center}
\includegraphics[width=0.49\textwidth]{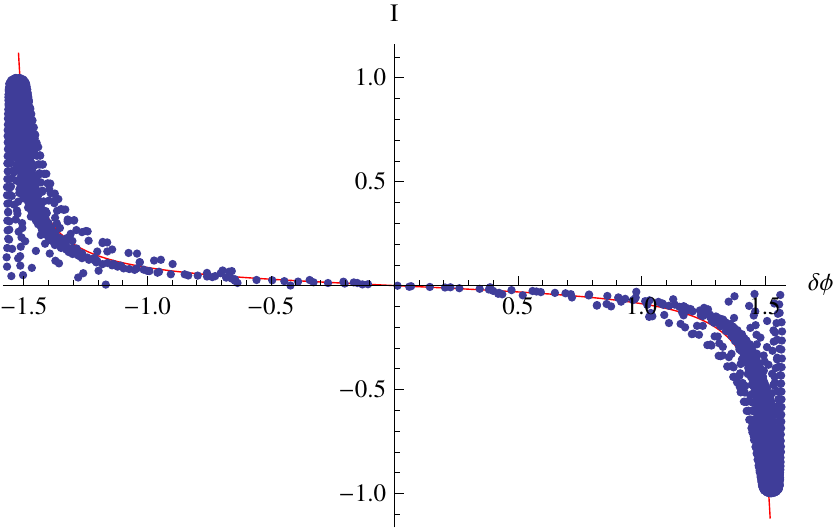}
\includegraphics[width=0.49\textwidth]{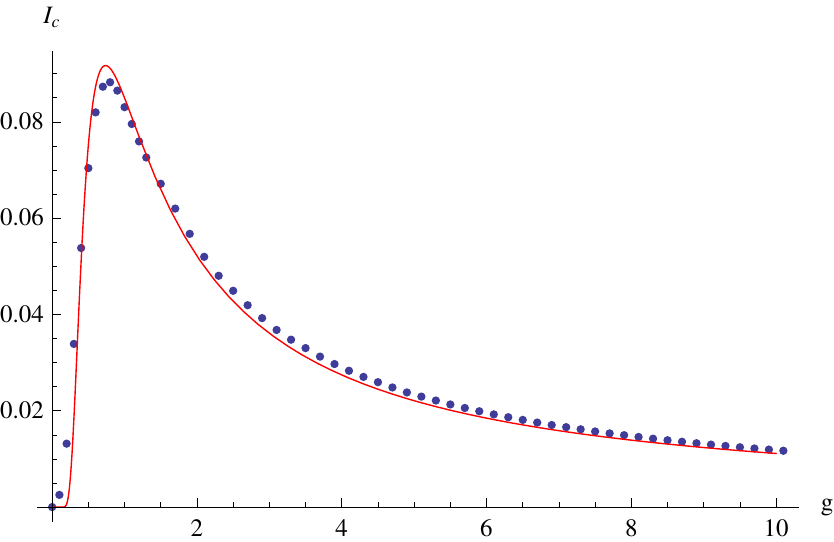}
\caption{Left: fit for the $I$-$\delta \phi$ characteristics with 
$g=2.3$, $\lambda=0.05$, $D=1$, $\delta M(t=0)=1$, $N=8$, $M_{T}=7$ - 
blue circles are numerical 
points obtained from the quantum dynamics, and the 
red line is the fit according 
Eq. (\ref{Itan}). 
Right: critical current fit -  the blue circles are numerical 
results, while the red line is Eq. (\ref{Ic-fit}) 
with $c \simeq 0.27$.}
\label{fig:I_c}
\end{center}
\end{figure}

As discussed in the Appendix \ref{appendix:dyn}, for the linear two-state 
model the current is proportional to the tangent of the phase difference: 
$I \propto \tan{\delta\phi}$ [see Eq. (\ref{tangent})]. 
The current-phase characteristic 
can be therefore written as
\begin{equation}\label{Itan}
I(\delta \phi) = I_c(g,\lambda) \tan{\delta\phi}
\end{equation}
and the critical current $I_c$ can be fitted from numerical data. An example 
is given in the left part of Figure \ref{fig:I_c}. 
We find that the critical current has a maximum around a finite value 
of $g$, as shown in the right part of Figure \ref{fig:I_c}: 
for the considered values of $N$ the maximum is at $g \simeq 1$, close to the 
unitary regime. $I_c$ can be fitted in the form
\begin{equation}\label{Ic-fit}
I_c(g,\lambda) = I_0 \lambda \frac{e^{-c/g^2}}{g}:
\end{equation}
$I_0$ depends mostly on $N$. Notice that the relation 
(\ref{Ic-fit}) has a maximum at $g^*=\sqrt{2c}$. For the parameters 
of Figure \ref{fig:I_c} we find  
$c \simeq 0.27$, nearly independent on $\lambda$.

We stress that the fit needed to identify the critical current is done 
using the linear two--mode model: the validity of the fit relies on the fact 
the two lowest levels are the ones mainly involved in the dynamics, which 
is the case for small imbalances ($D=1$). Deviations are observed 
for larger values of $D$, as we are going to discuss.

\begin{figure}
\begin{center}
\includegraphics[width=0.49\textwidth]{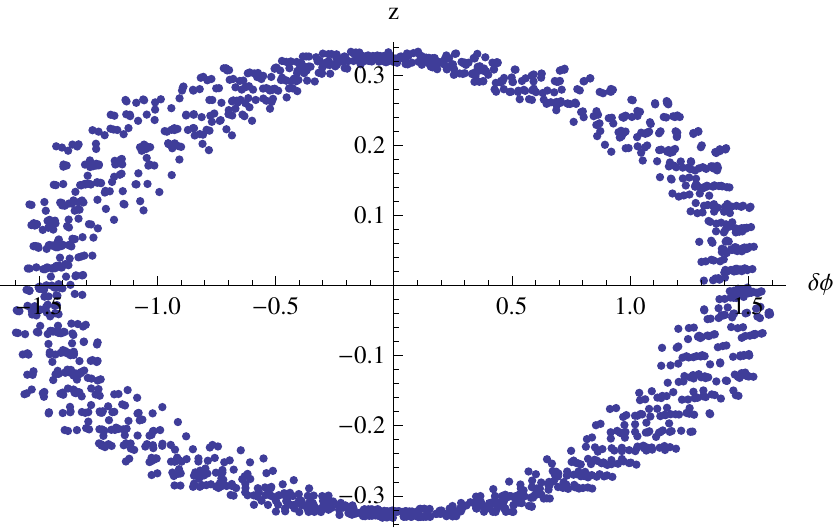}
\caption{Number-phase diagram for $N=8$, $D=2$, $\delta M(t=0)=2$, 
$M_T=6$, $\lambda=0.05$ and $g=0.4$.}
\label{fig:deltaM_2}
\end{center}
\end{figure}

It is an interesting issue to explore what happens when more levels, 
inserted in a band structure as the one described in Section 
\ref{sec:spectrum}, participate 
the dynamics: with $D=2$ and $\delta M(t=0)=2$ 
the phase diagram shows a typical ellipsoid form. 
An example of number-phase portrait is given in 
Figure \ref{fig:deltaM_2}. We see that 
the phase range depends only on the interaction, 
while the amplitude of the population oscillations depends 
on the initial relative phase given to the system through (\ref{initialstate}).

\begin{figure}
\begin{center}
\includegraphics[width=0.49\textwidth]{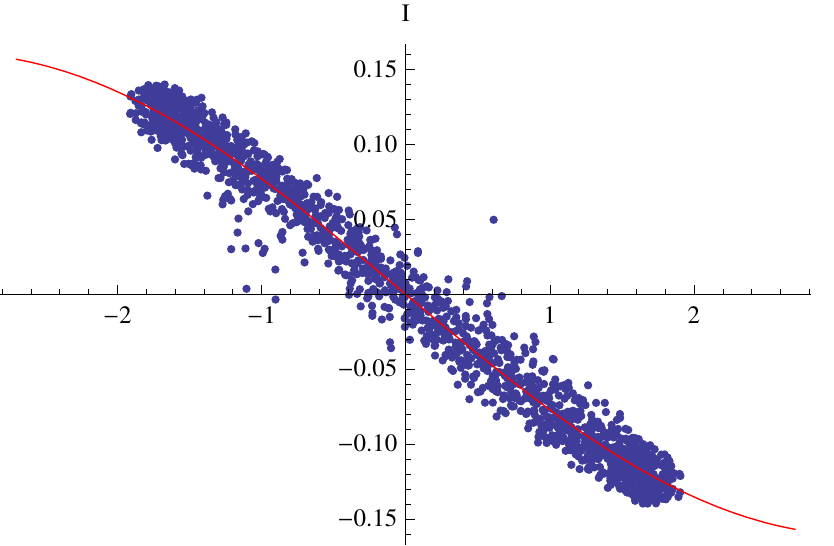}
\includegraphics[width=0.49\textwidth]{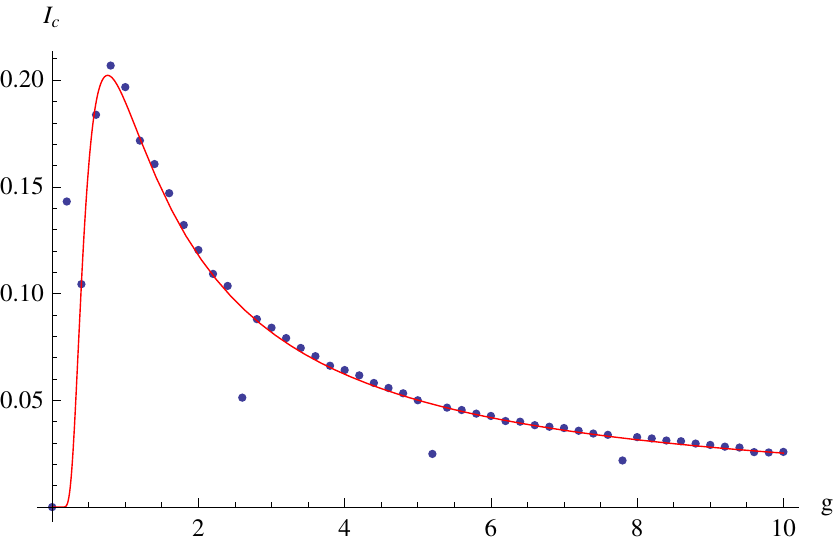}
\caption{Left: current-phase characteristics for $g=3.8$, $\lambda=0.05$, 
$D=2$, $\delta M(t=0)=2$, $N=8$, $M_{T}=7$ -   blue circles are numerical 
points obtained from the quantum dynamics, and the 
red line is the fit according 
Eq. (\ref{Isin}). 
Right: critical current vs. $g$ - the blue circles are 
the numerical results, while the red line is Eq. (\ref{Ic-fit}).}
\label{fig:fit_sin}
\end{center}
\end{figure}

The numerical study of the current phase characteristics reveals that 
for $D=2$ the relation (\ref{Itan}) does not provide a good way of fitting 
the critical current: the numerical results are plotted in the 
left part of Figure \ref{fig:fit_sin}. 
We find that a good approximation of the current-phase 
characteristics is given by 
\begin{equation}\label{Isin}
I(\delta\phi) = I_c(g,\lambda) \sin\frac{\delta\phi}{2}
\end{equation}
with $I_c$ given by (\ref{Ic-fit}) \cite{note1}, as it can be seen 
in the right part of Figure \ref{fig:fit_sin}.
We observe that such a dependence for the current-phase characteristics 
was found for a weak, point-like barrier in the WKB approximation 
in the Bogoliubov-de Gennes equation \cite{BH91}. Since for large 
$N$ we expect a dependence $\propto \sin{\delta\phi}$ \cite{SPS10}, 
we attribute the result (\ref{Isin}) to the small $N$ considered: 
further numerical investigations with larger number of levels are needed 
in order to obtain the current-phase characteristic for intermediate and large 
$N$ for the coupled Richardson models.

A check of Eq. (\ref{Isin}) and of the data presented in Figure 
\ref{fig:fit_sin} can be obtained by doing the Fourier transform 
of $\delta M(t)$ with $D=2$: as a function of $g$, to a very good approximation 
the dominant frequency of $\delta M(t)$ (i.e., 
the Fourier component with the highest weight)  
turns out to be proportional to the critical current given 
by (\ref{Ic-fit}).


An important prediction of the nonlinear two-state model 
is that there is a critical 
initial imbalance for which self-trapping occurs \cite{SFGS97}: given 
the limitation on the maximum value of $D$, we cannot explore larger 
initial imbalances. What is observed, instead, is that the 
amplitude of the fastest oscillations of $\delta M(t)$ is increased and that 
the period of the slowest ones is decreased more and more, as $1/g$. 
The time period of both $\delta M(t)$ and $\delta \phi(t)$ become larger 
and the oscillations exhibited by $\delta \phi(t)$ (as the ones 
seen in the left part of Figure \ref{fig:phase}) become as well larger. 
The scenario is that of a large crossover to a confined regime, 
in which the occupation oscillations have infinite period at $g$ very large. 
This may be a finite-$N$ effect, and one could expect that this 
eventually leads to a transition in the thermodynamic limit. 

The initial phase can also be varied with initial imbalance 
$\delta M(t=0)=2$. It is interesting to note that the for most of 
the values of $g$, the phase runs: nevertheless, the time evolution of 
the mean phase locks it around some large-period oscillations.  
We conclude by 
observing that similar results are found decreasing the coupling $\lambda$: 
further investigations to study self-trapping effects at very small values 
of the coupling are needed. An analysis of larger imbalances and larger 
$N$ (eventually with very small coupling) 
is therefore needed to study self-trapping effects, and 
more in general non-linear effects, through the crossover.

\section{Conclusions}\label{sec:concl}
We have studied the emergence of a definite relative phase between 
ultrasmall metallic grains (and in general finite-size systems 
of attractively interacting fermions) modeled by weakly 
coupled Richardson models. We have introduced and discussed 
a way of extracting the relative phase and its variance
 from the many-body wavefunction, in order to precisely quantify 
whether a definite relative phase emerges. 

We have also related the coherent behavior to the 
spectrum of the coupled systems
and suggested a criterion to characterize 
the crossover between the BCS and BEC regimes, showing  
that these regimes are clearly distinguishable 
by the spectrum of the coupled models.

Moreover, we have performed a numerical analysis of the exact dynamics 
of the two weakly coupled Richardson Hamiltonians, after a weak 
tunneling term is turned on. 
We used a linear superposition of the eigenstates of the two uncoupled systems,
with a different number of pairs ($D$ being such difference), 
as initial states: these states are then evolved according to the full 
Hamiltonian including the tunneling Hamiltonian, weakly 
coupling the two systems. 
We found that a definite relative phase difference emerges even for a small numbers of pairs ($\sim 8-10$). 
Therefore, the current-phase characteristics 
could be obtained for values of the bare pairing strength for which 
the equilibrium properties of the uncoupled models are well 
approximated by the BCS theory. 
We showed that, for small initial imbalances ($D=1$), a 
two-state model gives a reasonably good description of the dynamics and of 
the current-phase characteristics.

Finally, we have presented the critical current as a function 
of the pairing parameter, finding that it has a maximum 
around the unitary regime, even with 
a number of pairs $\sim 8$. The phase portrait 
was studied for small initial imbalances ($D \le 2$). 

The requirement of having a definite phase difference 
among the two systems with a limited total number of pairs ($\le 10$) 
prevented us to analyze values of the initial population imbalance ($D>2$): 
for these large initial imbalances the relative 
phase is well defined only for very strong pairing interaction, well beyond the 
unitary limit and deep in the BEC regime. Further numerical 
investigations are required to consider larger sizes and larger 
initial imbalances (eventually with very small tunneling couplings), 
which may generate a definite relative phase across 
the BCS-BEC crossover: it is expected that a proper finite-size scaling 
may be crucial to identity non-linear self-trapping effects. 
We moreover regard as interesting the investigation 
of the effects on the relative phase of single-fermion tunneling terms: 
these terms might give a contribution on the BCS side of the crossover and 
produce a degradation of the relative phase, which should eventually form 
for larger sizes. Similarly, it would be stimulating to compare (eventually 
for larger systems) the results obtained from exact 
dynamics with the ones obtained using time-dependent mean-field approaches.

The rapid growth of the computational cost with the size of the systems 
represents a limitation on the total number of pairs as well: 
the Hilbert space could be further reduced in the strong 
coupling regime, yet not throughout the whole crossover. 
We conclude that it stands as an open issue, 
certainly deserving future work,
how our findings scale with the size of the system.


Our results can be applied to weakly coupled ultrasmall metallic 
grains and to cold atom experiments in which traps with few fermions are 
set at a distance that allows tunneling: the individuation 
of the relative phase between nearest neighboring sites makes possible 
in perspective to study Josephson dynamics and self-trapping 
systems also for larger imbalances, and to check the validity 
of two- and multi- mode ansatz.

We finally observe that in this paper we focused our attention to weakly 
coupled Richardson models, discussing the formation of a relative phase and 
the Josephson dynamics for a class of considered initial conditions. The 
extension of our method of defining a relative phase to the problem of 
the formation of a relative phase between general interacting 
(both integrable and non-integrable) mesoscopic systems 
could be relevant in a rather broad 
class of physical systems, including weakly coupled ultracold 
finite Bose gases, and it is in our opinion an interesting problem, worthwhile 
of future studies.

\acknowledgments
Discussions with G. Sierra, A. De Luca, T. Macrì,
A. Smerzi, L. Amico, R. Scott, L. Pitaevskii and S. Stringari are very
gratefully acknowledged. F.B. also thanks A. De Luca for 
collaboration on the implementation of the 
numerical solution of the Richardson equations.

\appendix

\section{Dynamical equations for the two-state model}\label{appendix:dyn}

A general description of the tunneling in superfluid/superconducting systems 
is provided by the Feynman two-state model \cite{FLS65}: the macroscopic 
wavefunctions $\psi_L$ and $\psi_R$ of 
the left and right systems obey the equations
\begin{eqnarray}\label{twostate}
&&i\hbar \frac{\partial \psi_L}{\partial t} = E_L \psi_L - K \psi_R  \\
&&i\hbar \frac{\partial \psi_R}{\partial t} = E_R \psi_R - K \psi_L. 
\label{twostate2}
\end{eqnarray}
The two-state model also describes also the tunneling of Bose-Einstein 
condensates in double well potentials \cite{SFGS97}: the effect of 
the interactions between atoms in the wells results in cubic terms 
of the form $U \left| \psi_L \right|^2 \psi_L$ 
and $U \left| \psi_R \right|^2 \psi_R$ added to the right-hand sides 
of Eqs. (\ref{twostate})-(\ref{twostate2}). 
In our case, since the $D=1$ initial 
state (\ref{initialstate}) has mostly projections on the 
ground and first excited many-body states, we limit 
ourself to Eqs. (\ref{twostate})-(\ref{twostate2}) (with $U=0$).

Setting $\psi_{s}=\sqrt{M_{s}}e^{i\phi_{s}}$ (with $s=L,R$), 
the equations for $z \equiv (M_L-M_R)/(M_L+M_R)$ and 
$\phi \equiv \phi_R-\phi_L$ reads
\begin{eqnarray}\label{motion}
&&\hbar \frac{\partial z}{\partial t} = -2K \sqrt{1-z^2}\sin\phi,  \\
&&\hbar \frac{\partial \phi}{\partial t} = \frac{2Kz}{\sqrt{1-z^2}}\cos\phi 
\label{motion2}
\end{eqnarray}
for the symmetric case $E_L=E_R$.

The system (\ref{motion})-(\ref{motion2}) can be derived from 
the Hamiltonian 
\begin{equation}\label{PendulumHamiltonian}
{\cal H}= -2K\sqrt{1-z^2}\cos{\phi}
\end{equation}
in which the time evolution of the conjugated variables $\phi,z$ is 
found from
\begin{eqnarray}\label{motionfrompendulum}
&& \hbar \dot{z}=-\frac{\partial {\cal H}}{\partial \phi} \\
&& \hbar \dot{\phi} = \frac{\partial {\cal H}}{\partial z}.
\label{motionfrompendulum2}
\end{eqnarray}
By defining the angular variable $\theta$ 
such that $z=\sin\theta\,\in \left[-1,1\right]$, one finds 
from (\ref{PendulumHamiltonian})
\begin{eqnarray}\label{motion-theta}
&&\hbar \dot{\theta} = \mp 2 K \sin\phi  \\
&&\hbar \dot{\phi} = 2 K \tan\theta \cos\phi: 
\label{motion-theta-2}
\end{eqnarray}
where the $\mp$ sign accounts for the determination of the square root.
The time-dependent relative occupation is a function of time 
only through the relative phase $\phi$. 
Starting from (\ref{motion-theta})-(\ref{motion-theta-2}) and identifying with 
a prime the derivative with respect to $\phi$, one has
$$
\hbar \frac{d \theta}{d t}=\hbar \frac{d \theta}{d \phi}\frac{d \phi}{d t} = 
\mp 2K\sin \phi,
$$
from which
$$
\tan\theta \frac{d \theta}{d \phi} = \mp \tan \phi.
$$
By integration one obtains
\begin{equation}\label{thetaofphi}
\cos\theta = \frac{A_0}{\cos\phi},
\end{equation}
where the constant $A_0= \mp \cos\phi_0 \cos\theta_0$ is 
fixed by the initial conditions. 
Defining the current $I$ as $I=\dot{M}_L$ one has 
$I=M_T \dot{z}/2$ where $M_T=M_L+M_R$ is the total number 
of particles (pairs, in our case). Using 
(\ref{motion}) one has 
\begin{equation}\label{tangent}
I(\phi)=\frac{M_T}{2} \dot{\theta} \cos \theta = 
-\frac{K M_T A_0}{\hbar} \tan \phi.
\end{equation}
We conclude the Appendix by observing that for the linear two-state model 
here considered ($U=0$) the phase difference 
does not overcome the value $\pi/2$ (more precisely, if 
$\left| \phi(t=0) \right|<\pi/2$, then $\left| \phi(t) \right|<\pi/2$).


\begin{thebibliography}{XX}
\bibitem{note1}
Fitting the current-phase characteristics via the function 
$I=I_c \sin{A\delta\phi}$ we obtain $A=0.5 \pm 0.1$ and $I_c$ well 
approximated by (\ref{Ic-fit}).
\end{thebibliography}

\end{document}